\documentclass[10pt,fleqn,a4paper,twoside]{article}
\usepackage{graphicx}
\usepackage{amsmath}
\usepackage[version=4]{mhchem}
\usepackage{siunitx}
\usepackage{tabularx}
\usepackage{longtable}
\usepackage{float}
\usepackage{placeins} 
\usepackage{xcolor}
\usepackage{subfigure}
\usepackage{amsmath}
\usepackage[sort&compress,numbers]{natbib}
\usepackage{hyperref}
\usepackage[margin=1in,total={8in,10in}]{geometry}

\begin{document}

\title{Development of Transonic Unsteady Aerodynamic Reduced-Order Models Using System Identification Techniques}
\author{Ana Cristina Neves Carloni\footnote{Instituto Tecnológico de Aeronáutica, 12228-900 São José dos Campos, SP, Brazil} \and João Luiz F. Azevedo\footnote{Instituto de Aeronáutica e Espaço, 12228-904 São José dos Campos, SP, Brazil}}
\date{}
\vspace{-.5cm}
\maketitle

\begin{abstract}
The present paper develops a reduced-order model capable of modeling unsteady aerodynamic loads in the transonic regime using system identification techniques. The computational fluid dynamics (CFD) calculations are based on the Euler equations and the code uses a finite volume formulation for general unstructured grids. A centered spatial discretization with added artificial dissipation is used, and an explicit Runge-Kutta time marching method is employed. For comparison reasons, unsteady calculations are performed using mode-by-mode and simultaneous excitation approaches. System identification techniques are employed to allow the splitting of the aerodynamic coefficient time histories into the contribution of each individual mode to the corresponding aerodynamic transfer function. Such methodology is applied to model the aerodynamic terms of the aeroelastic state-space system particularly of a NACA 0012 airfoil-based typical section. Results demonstrate the importance of signal processing techniques to compute the aerodynamic transfer functions and also the advantageous applicability of transonic unsteady aerodynamic reduced-order models to perform aeroelastic analyses in the frequency domain.\end{abstract}
\vspace{.2cm} \hspace{.75cm}

\section{Introduction}

Aeroelastic phenomena arise when structural deformations induce additional aerodynamic forces. These additional aerodynamic forces, in turn, can produce additional structural deformations which will induce even greater aerodynamic forces. These interactions may decrease until the stability condition is reached, or they may diverge and cause catastrophic structural failure \citep{bisplinghoff1955aeroelasticity,wright2008introduction}. In recent years, modern aircraft designs are evolving towards including large aspect ratio, increased flexibility, and additional complexity \citep{waite2019reduced,skujins2014reduced}. These features certainly intensify the structural flexibility, which is the main driver for the occurence of several aeroelastic phenomena \citep{bisplinghoff1955aeroelasticity}. This raises the need to develop an accurate and affordable simulation process for addressing transonic aeroelastic systems. Traditional aeroelastic analysis models the fluid-structure interation by explicitly implementing an interactive numerical process between the aerodynamic and dynamic-structural systems \citep{silva2009development,camilo2013hopf,silva2004development,silva2008simultaneous,silva2014evaluation}. Consistently, any parametric or flight condition variation necessarily demand a repetitive use of high-fidelity CFD codes. Given the available computational power, the traditional aeroelastic approach is prohibitive for engineering applications that require an extensive sweep over the entire flight envelope. One of the most prominent approaches to tackle this challenge is the reduced-order model (ROM) formulation, where the prevailing dynamics of the aeroelastic system is captured. The present effort builds upon previous work \citep{marques2008numerical,azevedo2012efficient,marques2008z,azevedo2013effects} that takes advantage of the reduced-order model formulation in transonic aeroelastic applications. Essentially, the goal of the present work is to develop transonic unsteady reduced-order model using system identification techniques in order to consistently apply the simultaneous excitation approach. Such approach ultimately enables the aeroelastic stability analysis to be performed with a single unsteady CFD run.

The CFD code used here is based on the 2-D Euler equations, which are discretized using a finite volume approach for unstructured grids. A centered scheme with added artificial dissipation is used for spatial discretization and explicit Runge-Kutta methods are employed for time marching. The present paper concentrates initially in identifying the aerodynamic transfer functions from a step excitation in each natural mode of the NACA 0012 airfoil-based typical section. This system identification procedure is based on the computation of power spectral densities of the inputs and outputs of the dynamic system. A relatively simple approach to estimate the transfer functions is by a division in the frequency domain for uncorrelated input signals and their derivatives. The effects of using different signal processing configurations to identify the transfer functions are studied. Once the transfer functions are obtained, they can be estimated by a rational-function approximation (RFA) in the Laplace domain \citep{tiffany1987nonlinear,eversman1991consistent} in order to better suit the solution of the aeroelastic eigenvalue problem. Finally, the solution of such eigenvalue problem for varying dynamic pressures yields a root locus from which one can estimate the flutter boundary of the configuration.

\section{Theoretical Formulation}

\subsection{Aeroelastic System}

\begin{figure}[hbt!]
\centering
\includegraphics[width=.45\textwidth]{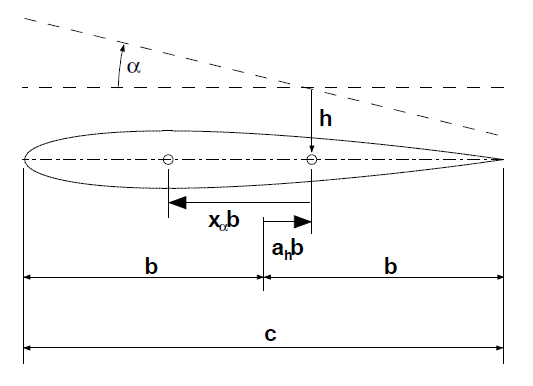}
\caption{\label{fig:figure1}Typical section configuration.}
\end{figure}

The structural model considered in the present work is the typical section, which is widely known and reported in literature \citep{bisplinghoff1955aeroelasticity,azevedo2013effects,marques2008z}. The dynamic system represented in the typical section is a rigid airfoil with two degrees of freedom, plunge and pitch. As presented in Fig.\ \ref{fig:figure1}, $h$ is the vertical translation, positive downwards, and $\alpha$ is the pitch mode coordinate, positive in the nose-up direction. Besides, $c$ is the airfoil chord, $b$ is the semi-chord length, $x_{\alpha}$ is the distance from the elastic axis to the center of mass normalized by the semi-chord, and $a_{h}$ is the distance from mid-chord to the elastic axis normalized by the semi-chord. The computational mesh used here is the same from Azevedo \emph{et al.} \cite{azevedo2013effects}, which has about 16000 triangular control volumes and just over 290 points along the airfoil. The considered mesh movement takes into account the body motions involved in unsteady calculations by rigidly moving the mesh accordingly. In this approach, the far-field boundary conditions are adjusted to account for the boundary movement together with the rest of the mesh, which is essentially displaced as a rigid body in order to follow the motion of the airfoil.

For aeroelastic problems, it is very common to represent the aerodynamic effects exclusively through the resulting forces and moments acting on the structure as a forcing term so that the governing equation for this dynamic system is given by
\begin{equation}
    \label{eq:equation1}
    [M] \{\Ddot{\eta}(t)\} + [K]\{\eta(t)\} = \{Q_{a}(t)\}
\end{equation}
\noindent where the generalized mass and stiffness matrices for the typical section model are, respectively,
\begin{equation}
    [M] = \begin{bmatrix}
            1 & x_{\alpha} \\
            x_{\alpha} & r_{\alpha}^2 
          \end{bmatrix}
\quad \quad \mbox{ and } \quad \quad 
    [K] = \begin{bmatrix}
            \omega_{h}^2 & 0 \\
            0 & r_{\alpha}^2 \omega_{\alpha}^2 
           \end{bmatrix} ,
\end{equation}
\noindent and the generalized coordinate and generalized aerodynamic forces are
\begin{equation}
    \{\eta(t)\} = \begin{Bmatrix}
                    \xi(t)  \\
                    \alpha(t) 
                  \end{Bmatrix}
\quad \quad \mbox{ and } \quad \quad 
    \{Q_{a}(t)\} = \begin{Bmatrix}
            \dfrac{Q_{a_h}(t)}{mb} \\
            \dfrac{Q_{a_\alpha}(t)}{mb^2}
          \end{Bmatrix} .
\end{equation}

In order to represent more general situations, it is convenient to nondimensionalize the aeroelastic equation. This is accomplished with the procedure proposed by Oliveira \cite{oliveira1993metodologia}, using a reference circular frequency to nondimensionalize the time variable,
\begin{equation}
    \Bar{t} = \omega_{r} t.
\end{equation}
\noindent With the application of the chain rule for the time derivatives, Eq.~(\ref{eq:equation1}) becomes
\begin{equation}
    [M] \{\Ddot{\eta}(\Bar{t})\} + [\Bar{K}]\{\eta(\Bar{t})\} = \{\Bar{Q}_{a}(\Bar{t})\}
\end{equation}
\noindent where
\begin{equation}
    [\Bar{K}] = \dfrac{1}{\omega_{r}^2} [K] 
    \quad \quad \mbox{ and } \quad \quad 
    \{\Bar{Q}_{a}(\Bar{t})\} = \dfrac{1}{\omega_{r}^2} \{Q_{a}(t)\} .
\end{equation}

The present work also attempts to efficiently determine the generalized aerodynamic force vector, $\{\Bar{Q}_{a}(\Bar{t})\}$, for an arbitrary structural behavior. However, it is very difficult to obtain a general expression for the aerodynamic response due to the nonlinearities of the aerodynamic equations. This problem is simplified by extending linearity concepts present in the formulation of the potential aerodynamic equations. As presented by Bisplinghoff \emph{et al.} \cite{bisplinghoff1955aeroelasticity}, the linear aerodynamic responses can be individually determined for each mode and then superimposed for more general responses. Based on these ideas, Oliveira \cite{oliveira1993metodologia} proposed the assumption of linearity of the aerodynamic response in the transonic regime regarding the modal motion. In addition, Marques and Azevedo \cite{marques2008z} show that there is a certain small amplitude range in which this hypothesis holds. For these reasons, the input signals applied to the aeroelastic system are restricted to small amplitude motions in the present work.

As a consequence of the linearity assumption with regard to the modal motion, it is possible to determine the aerodynamic response to a general structural behavior from the convolution of an impulsive or indicial aerodynamic solution \citep{bisplinghoff1955aeroelasticity,silva2004development}. The
convolution operation, however, is more easily handled in the frequency domain, in which it is represented by a simple multiplication operation. Hence, the aerodynamic forces due to the structural motion, linearized with regard to the modal displacements, can be written in the frequency domain. After some mathematical manipulations, one can define the generalized forces as
\begin{equation}
    \{\Bar{Q}_{a} (\kappa)\} = \dfrac{(U^{*})^2}{\pi \mu} [A(\kappa)] \{\eta(\kappa)\}
\end{equation}
\noindent where the mass ratio and the characteristic speed are, respectively,
\begin{equation}
    \mu = \dfrac{m}{\pi \rho_{\infty} b^2} 
    \quad \quad \mbox{ and } \quad \quad 
    U^{*} = \dfrac{U_{\infty}}{b \omega_r},
\end{equation}
\noindent and the aerodynamic influence coefficient matrix is given by
\begin{equation}
    [A(\kappa)] = \begin{bmatrix}
            -C_{l_{h}}(\kappa)/2 & -C_{l_{\alpha}}(\kappa) \\
             C_{m_{h}}(\kappa) & 2 C_{m_{\alpha}}(\kappa)
          \end{bmatrix} .
\end{equation}

\subsection{State-Space Analysis}

A state-space formulation is conveniently used to describe a dynamic system in terms of first-order differential equations. In the present case, this formulation can be obtained by defining
\begin{equation}
\begin{aligned}
    \{x_{1}(\Bar{t})\} &= \{\eta(\Bar{t})\}, \\
    \{x_{2}(\Bar{t})\} &= \{\Dot{\eta}(\Bar{t})\} = \{\Dot{x}_{1}(\Bar{t})\}, \\
    \{x(\Bar{t})\} &= \begin{Bmatrix} 
                        \{x_{2}(\Bar{t})\} \\
                        \{x_{1}(\Bar{t})\}
                       \end{Bmatrix} .
\end{aligned}
\end{equation}
\noindent Hence, the governing equation of motion becomes
\begin{equation}
    \label{eq:equation12}
    [\Tilde{M}] \{\Dot{x}(\Bar{t})\} + [\Tilde{K}]\{x(\Bar{t})\} = \{\Tilde{q}(\Bar{t})\},
\end{equation}
\noindent where
\begin{equation}
    [\Tilde{M}] = \begin{bmatrix}
            [M] & [0_{2 \times 2}] \\
            [0_{2 \times 2}] & [I_{2 \times 2}] 
          \end{bmatrix} ,
\quad \quad
    [\Tilde{K}] = \begin{bmatrix}
            [0_{2 \times 2}] & [\Bar{K}] \\
            -[I_{2 \times 2}] & [0_{2 \times 2}] 
           \end{bmatrix} ,
\quad \quad \mbox{ and } \quad \quad           
   \{\Tilde{q}(\Bar{t})\} = \begin{Bmatrix}
                            \{\Bar{Q}_{a}(\Bar{t})\} \\
                            \{0_{2 \times 1}\}
  \end{Bmatrix} .
\end{equation}

As one can represent the aerodynamic forces in the frequency domain, the aeroelastic system can be more easily studied in the Laplace domain. Applying the Laplace transform to Eq.~(\ref{eq:equation12}), one obtains
\begin{equation}
    \Bar{s} [\Tilde{M}] \{X(\Bar{s})\} + [\Tilde{K}]\{X(\Bar{s})\} = \{\Tilde{Q}(\Bar{s})\},
\end{equation}
\noindent where
\begin{equation}
    \Bar{s} = \dfrac{s}{\omega_r}
\end{equation}
\noindent is the dimensionless Laplace transform complex variable. The idea behind this procedure is to evaluate the aerodynamic influence coefficient matrix over a reduced frequency range of interest and, by making use of the analytical continuation principle, to extend such result to the entire complex plane.

Regardless of which polynomial is used to approximate the aerodynamic responses in the Laplace domain, the resulting state-space system of the typical section is always described as
\begin{equation}
    \{\Dot{\chi}(\Bar{t})\} = [D] \{\chi(\Bar{t})\},
\end{equation}
\noindent where $\{\chi(\Bar{t})\}$ is the new state vector that includes the aerodynamic state variables. A detailed formulation of the stability matrix of the system, $[D]$, can be found in literature \citep{eversman1991consistent,azevedo2013effects,marques2008z}. Finally, the aeroelastic stability analysis can then be reduced to the classical eigenvalue problem for each value of the characteristic speed parameter, $U^{*}$, such that
\begin{equation}
    ([D] - \Bar{s} [I_{N \times N}]) \{\chi(\Bar{s})\} = \{0_{N \times 1}\}
\end{equation}
\noindent where $N=2 n_{a}+4$ for an aeroelastic system with two structural degrees of freedom and $n_{a}$ aerodynamic states.

\subsection{Eversman and Tewari Polynomials}

In computational procedures, the aerodynamic responses in the frequency domain consist of sets of numerical discrete values, which are not convenient for the solution of Eq.~(\ref{eq:equation12}). Hence, it is necessary to approximate these data using interpolating polynomials. There are some possible polynomials reported in the literature \citep{tiffany1987nonlinear,eversman1991consistent}. The interpolating polynomial proposed by Eversman and Tewari \cite{eversman1991consistent}, particularly the one that does not include any provision for the treatment of repeated or very close poles, is initially selected by the authors for implementation and tests. The choice of rational-function approximation (RFA) is based on the fact that such polynomials are the most commonly used in similar applications, and that they are conveniently constructed for the formulation of aerodynamic state variables. In the Laplace domain, this rational-function approximation is written as
\begin{equation}
    \label{eq:equation16}
    [A(\Bar{s})] = [A_{0}] + [A_{1}] \dfrac{\Bar{s}}{U^{*}} + [A_{2}] \left( \dfrac{\Bar{s}}{U^{*}} \right)^{2} + \sum_{n=1}^{n_{\beta}} \left( [A_{n+2}] \dfrac{U^{*}}{\Bar{s}+U^{*} \beta_{n}}  \right),
\end{equation}
\noindent where the $\beta_{n}$'s introduce the aerodynamic lags with respect to the structural modes. The variables of the polynomial are determined by an optimized least squares approximation method, and the number of augmented states equals the number of added poles, \textit{i.e.}, $n_{a}=n_{\beta}$. The formulation presented in Eq.~(\ref{eq:equation16}) is referred to as the first form of the Eversman and Tewari polynomials.

\section{Results and Discussion}

\subsection{Simulation Procedure}

A NACA 0012 airfoil-based typical section at $M_{\infty}=0.8$, \textit{i.e.}, in the transonic regime, and $\alpha_{0}=0$ is considered throughout this paper. The structural parameters are $a_h=-2.0$, $x_{\alpha}=1.8$, $r_\alpha=1.865$, $\omega_h=\omega_\alpha=100$ rad/s, $\mu=60$, and $\omega_r=\omega_\alpha$ is used as a reference value. The unsteady CFD run is initialized with a steady-state solution, which is obtained by the same CFD code running in steady Euler mode with a variable time step method. From the converged stationary solution, the mesh is rigidly moved in a prescribed pattern and a total of 100,000 time steps of unsteady flow are computed with a constant $\Delta \Bar{t}$ of 0.003 dimensionless time units. 

\begin{figure}[hbt!]
    \begin{center}
    	 \subfigure[First set of Walsh function inputs (WF1).]{ \includegraphics[scale=0.32,trim = 1.5cm 6cm 2cm 7cm,clip]{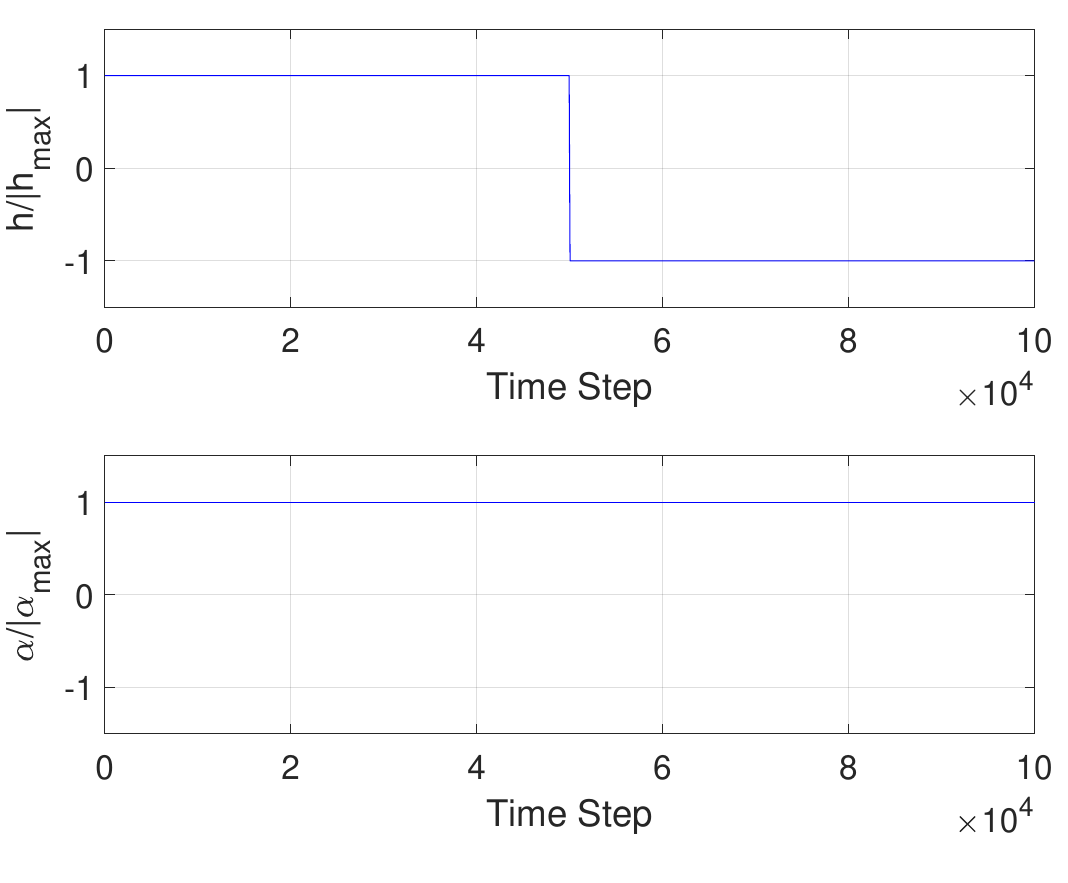}} \quad
    	 \subfigure[Second set of Walsh function inputs (WF2).]{ \includegraphics[scale=0.32,trim = 1.5cm 6cm 2cm 7cm,clip]{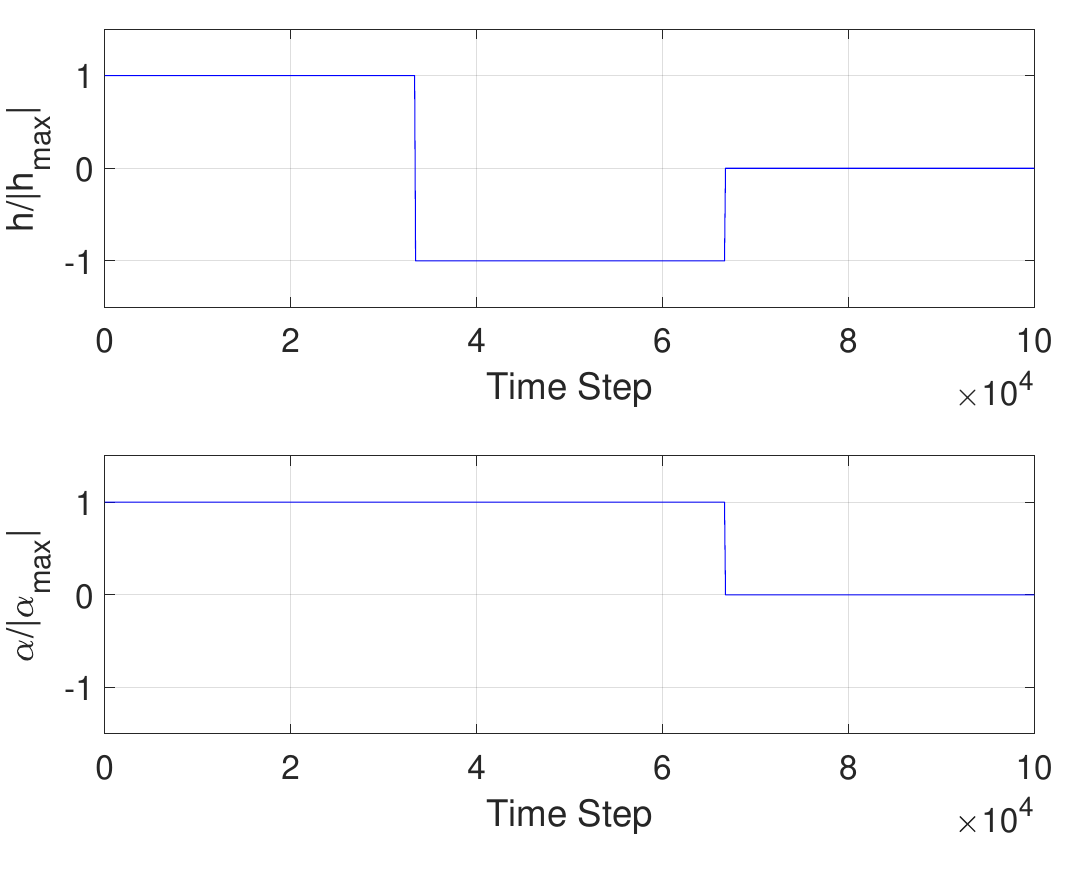}} \quad
    	 \subfigure[Third set of Walsh function inputs (WF3).]{ \includegraphics[scale=0.32,trim = 1.5cm 6cm 2cm 7cm,clip]{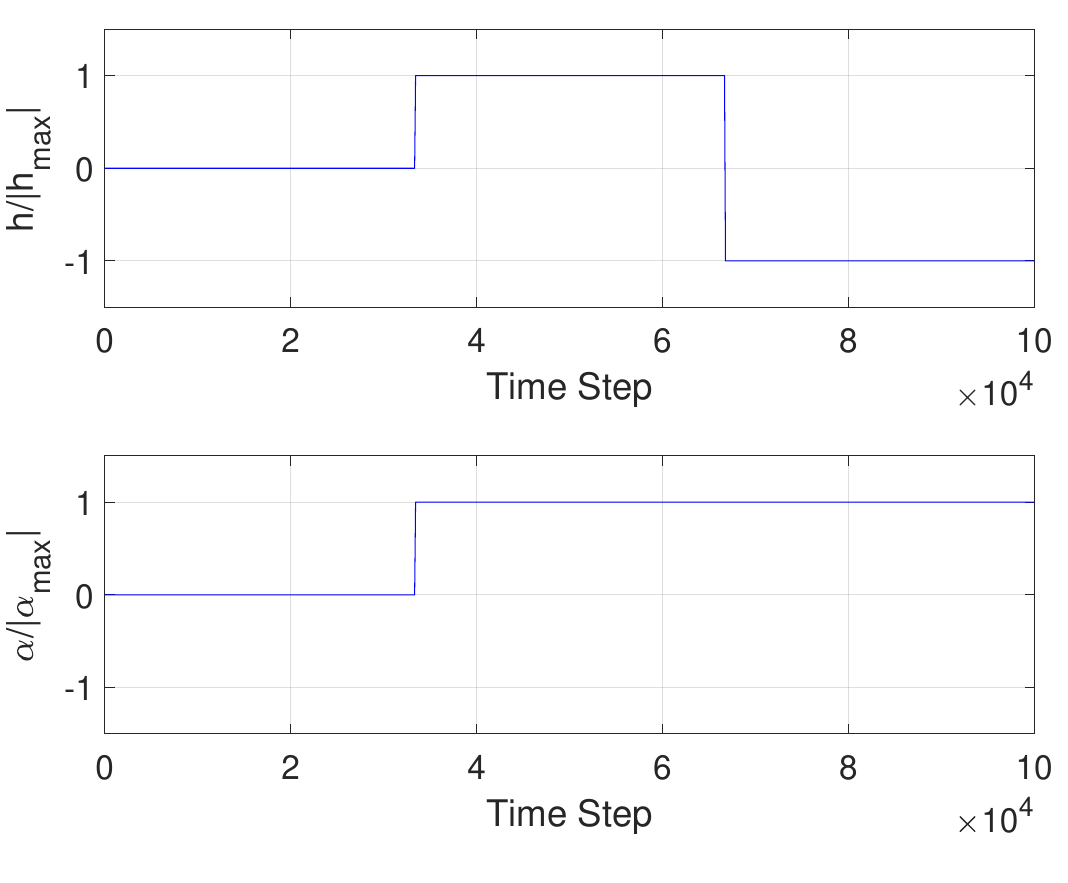}} \quad
    	\subfigure[Fourth set of Walsh function inputs (WF4).]{ \includegraphics[scale=0.32,trim = 1.5cm 6cm 2cm 7cm,clip]{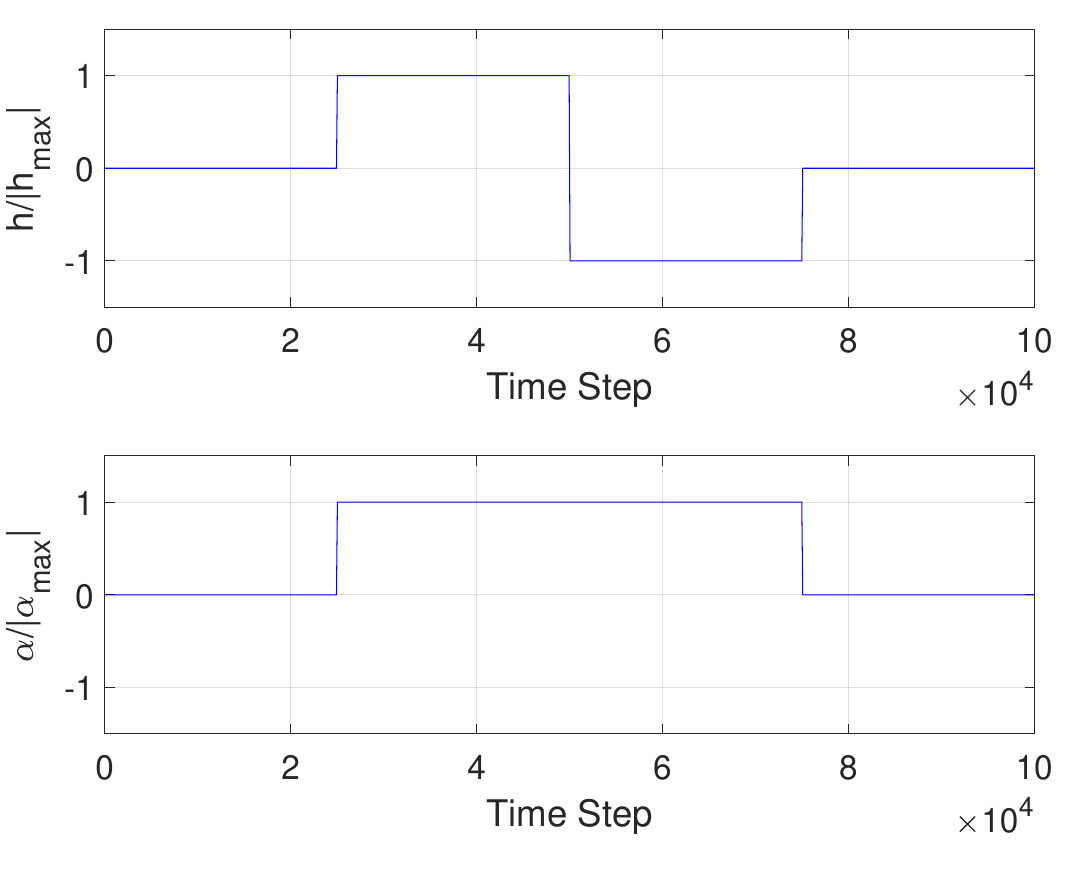}} \quad
    	 \subfigure[Fifth set of Walsh function inputs (WF5).]{ \includegraphics[scale=0.32,trim = 1.5cm 6cm 2cm 7cm,clip]{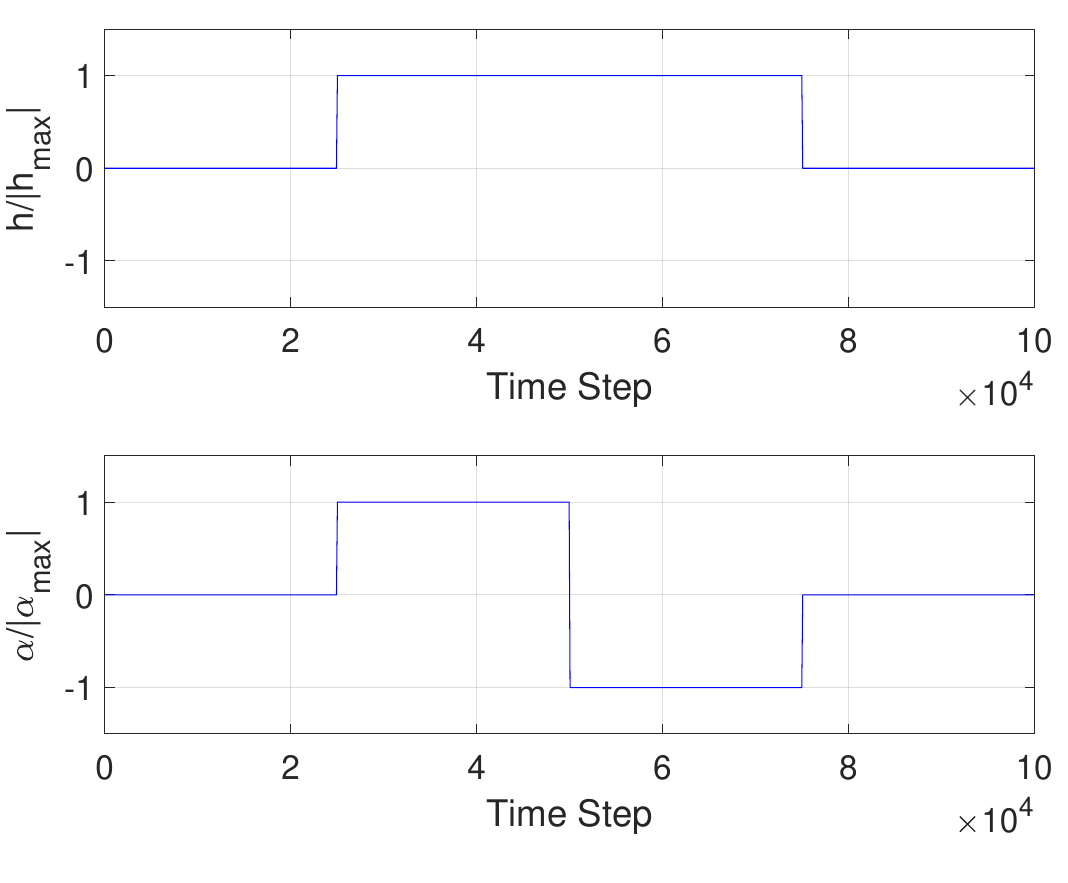}} 
    	 \caption{\label{fig:figure2}Normalized motions prescribed by different orthogonal inputs.}
    	 \vspace{-0.7cm}
    \end{center}
\end{figure}

The prescribed airfoil movements are defined as step inputs and Walsh functions. The Walsh functions are a set of block step functions that form an orthogonal basis of the square-integrable functions. Each function takes positive and negative unitary values and each step block period is an integer multiple of a $2^n$ division of the function length. Each of these functions may be seen as a line, or a column, of a Hadamard matrix of order $2^n$ \citep{hadamard93}. This family of functions is similar to step inputs and, therefore, embodies their impulsive nature with regard to frequency bandwidth. It is important to emphasize that the step inputs are used in a mode-by-mode fashion whereas the Walsh functions are used to simultaneously excite all the system modes. Furthermore, the maximum amplitudes considered here are 0.000001$c$ for the plunging mode and 0.0001 deg. for the pitching degree of freedom. The reason for choosing such low amplitudes, as discussed earlier, is to remain within the ``quasi-linear'' region around the nonlinear steady solution and to allow the CFD code to accurately propagate the disturbances from the discrete motion.

Moreover, the five cases considered for the Walsh inputs are illustrated in Fig.\ \ref{fig:figure2}. Many configurations have been studied by the authors with sets of different sizes of zero-motion regions and locations. These zero-motion regions attempts to guarantee that the CFD computation starts and ends with a solution that does not have any perturbations from the prescribed motion. The choice of these sets is based on the work of Azevedo \emph{et al.} \cite{azevedo2013effects}. An advantage of the proposed procedure is that the simulations with any set of Walsh functions require only one unsteady CFD run, whereas those with a discrete step would require two simulations in the present case, \textit{i.e.}, one for each modal movement. The solutions provided by the CFD code are then used by a system identification routine developed in MATLAB$^{\copyright}$. Both mode-by-mode and simultaneous simulations are analyzed with the power spectral density method for comparison reasons. The transfer functions obtained in this way are compared to those presented in Refs.\ \cite{marques2008numerical} and \cite{azevedo2013effects}. Lastly, the discrete values referring to each transfer function are interpolated using the first form of the Eversman and Tewari polynomials in order to provide a suitable input for flutter analysis.

\subsection{Aerodynamic Transfer Functions}

The validation of the power spectral density procedure was achieved by comparing the transfer functions obtained with a single unsteady CFD run with the ones previously calculated using a mode-by-mode approach as shown in Ref.\ \cite{azevedo2012efficient}. Besides, when transforming the discrete aerodynamic time histories into frequency domain transfer functions, one is faced with certain variables regarding the size of the sampled data and the number of overlapping points. To simplify the present analysis, the samples and overlaps were used in an integer value of information blocks. An information block is defined here by the length of the input discrete data where no variations occur. As a result, the five unsteady CFD results were, then, spread through a total of 18 different test cases. The number of information blocks, size of the sample, and the number of overlapping blocks adopted in each test case, as well as the input it refers to, are described in Table \ref{tab:table1}. The table also includes a test case, referred to as Case 00, that applies discrete step inputs in each natural mode separately. The present work initially applies a rectangular window with the same size of the sampled data, but further investigations conducted herein also analyze the effects of employing the Hanning window based on signal processing techniques. 

\begin{table}[hbt!]
    \caption{\label{tab:table1}System identification variables.} 
	\begin{center} {
		\begin{tabular}{ccccc}
		\hline
		Case & Inputs & Blocks & Sample & Overlap \\ \hline
        00   & Step  & 1      & 1      & 0       \\
        01   & WF1   & 2      & 2      & 0       \\
        02   & WF1   & 2      & 1      & 0       \\
        03   & WF2   & 3      & 3      & 0       \\
        04   & WF2   & 3      & 2      & 1       \\
        05   & WF2   & 3      & 1      & 0       \\
        06   & WF3   & 3      & 3      & 0       \\
        07   & WF3   & 3      & 2      & 1       \\
        08   & WF3   & 3      & 1      & 0       \\
        09   & WF4   & 4      & 4      & 0       \\
        10   & WF4   & 4      & 3      & 2       \\
        11   & WF4   & 4      & 2      & 1       \\
        12   & WF4   & 4      & 2      & 0       \\
        13   & WF4   & 4      & 1      & 0       \\
        14   & WF5   & 4      & 4      & 0       \\
        15   & WF5   & 4      & 3      & 2       \\
        16   & WF5   & 4      & 2      & 1       \\
        17   & WF5   & 4      & 2      & 0       \\
        18   & WF5   & 4      & 1      & 0       \\ \hline
	    \end{tabular}}
	    \vspace{-0.5cm}
	\end{center}
 \end{table}

\begin{figure}[hbt!]
    \begin{center}
    	 \subfigure[$G_{C_{l},h}$ results for Case 02 using rectangular window overlaid by Case 00 counterpart.]{ \includegraphics[scale=0.37,trim = 3.5cm 8.5cm 4.5cm 9cm,clip]{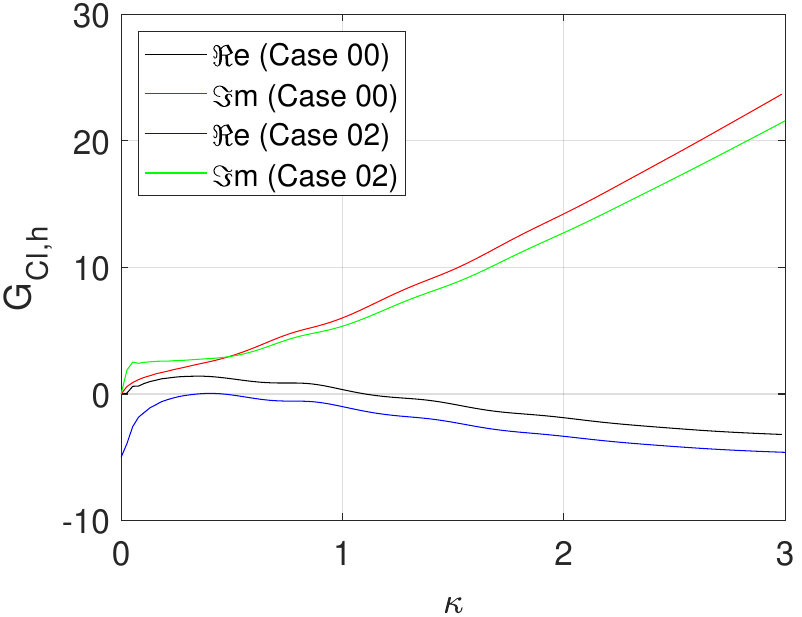}} \quad
    	 \subfigure[$G_{C_{l},\alpha}$ results for Case 02 using rectangular window overlaid by Case 00 counterpart.]{ \includegraphics[scale=0.37,trim = 3.5cm 8.5cm 4.5cm 9cm,clip]{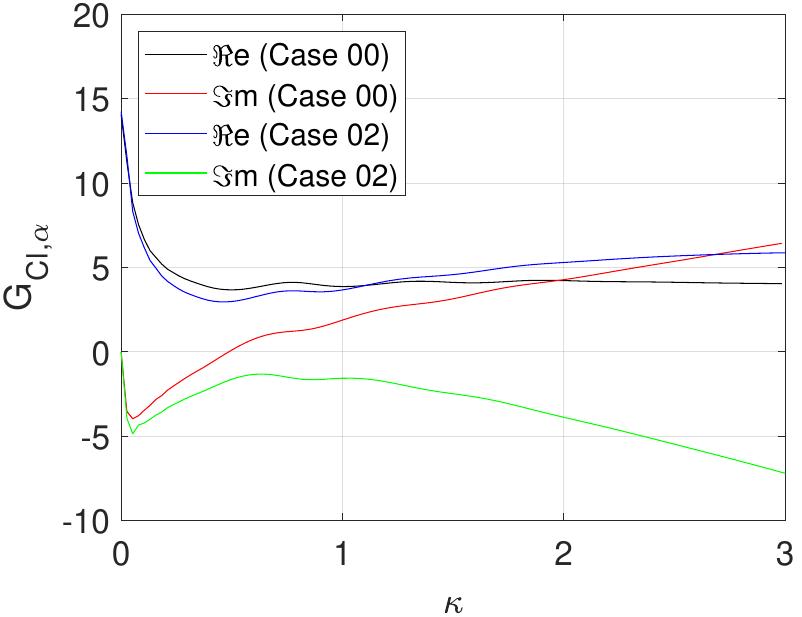}} \\
    	 \subfigure[$G_{C_{m},h}$ results for Case 02 using rectangular window overlaid by Case 00 counterpart.]{ \includegraphics[scale=0.37,trim = 3.5cm 8.5cm 4.5cm 9cm,clip]{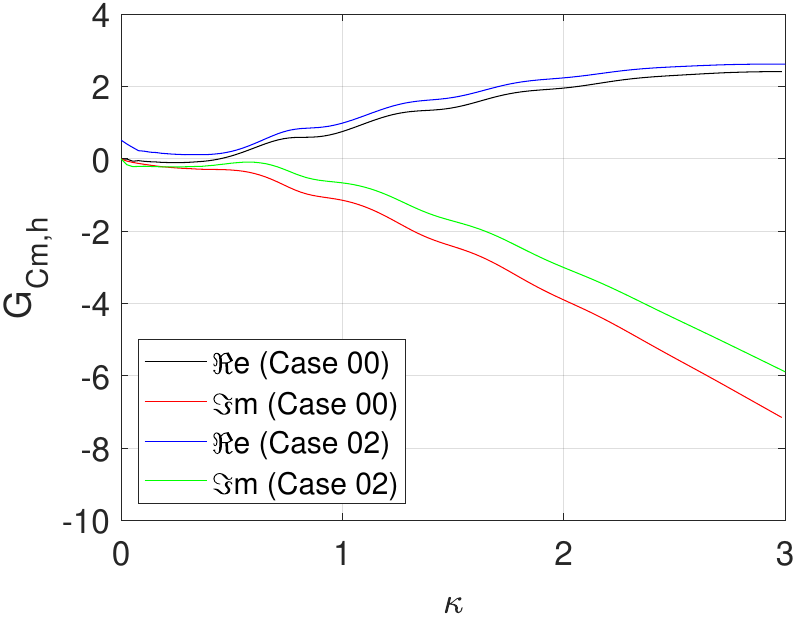}} \quad
    	\subfigure[$G_{C_{m},\alpha}$ results for Case 02 using rectangular window overlaid by Case 00 counterpart.]{ \includegraphics[scale=0.37,trim = 3.5cm 8.5cm 4.5cm 9cm,clip]{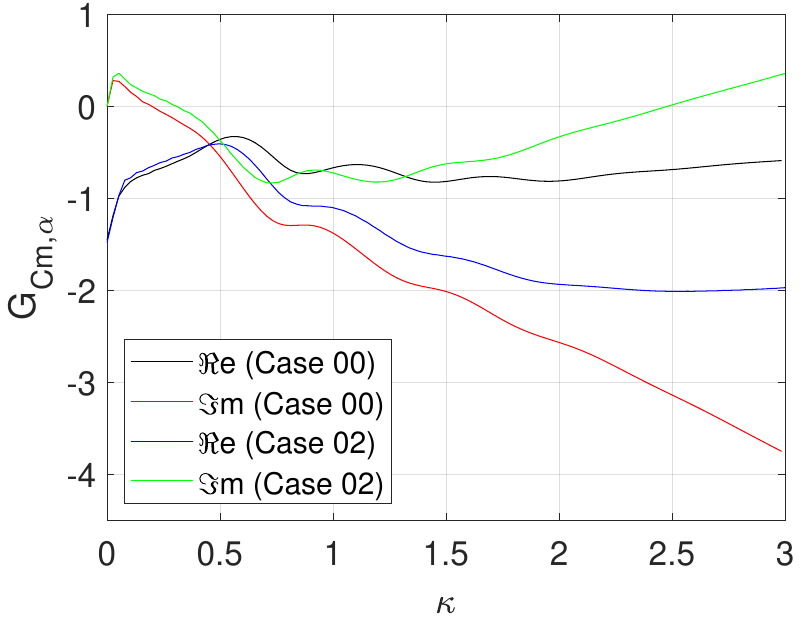}} 
    	 \caption{\label{fig:figure3}Aerodynamic transfer function for Case 02 using rectangular window.}
    	 \vspace{-0.2cm}
    \end{center}
\end{figure}

\begin{figure}[hbt!]
    \begin{center}
    	 \subfigure[$G_{C_{l},h}$ results for Case 08 using rectangular window overlaid by Case 00 counterpart.]{ \includegraphics[scale=0.37,trim = 3.5cm 8.5cm 4.5cm 9cm,clip]{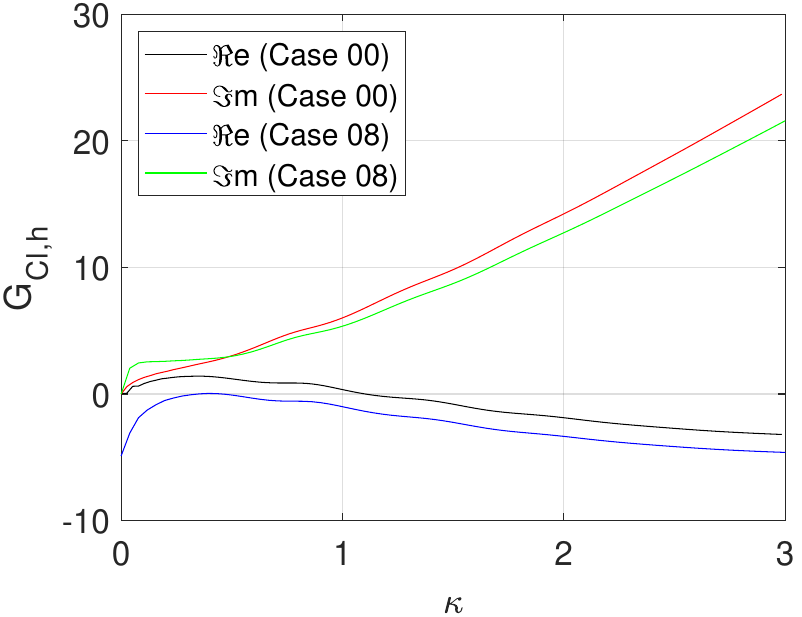}} \quad
    	 \subfigure[$G_{C_{l},\alpha}$ results for Case 08 using rectangular window overlaid by Case 00 counterpart.]{ \includegraphics[scale=0.37,trim = 3.5cm 8.5cm 4.5cm 9cm,clip]{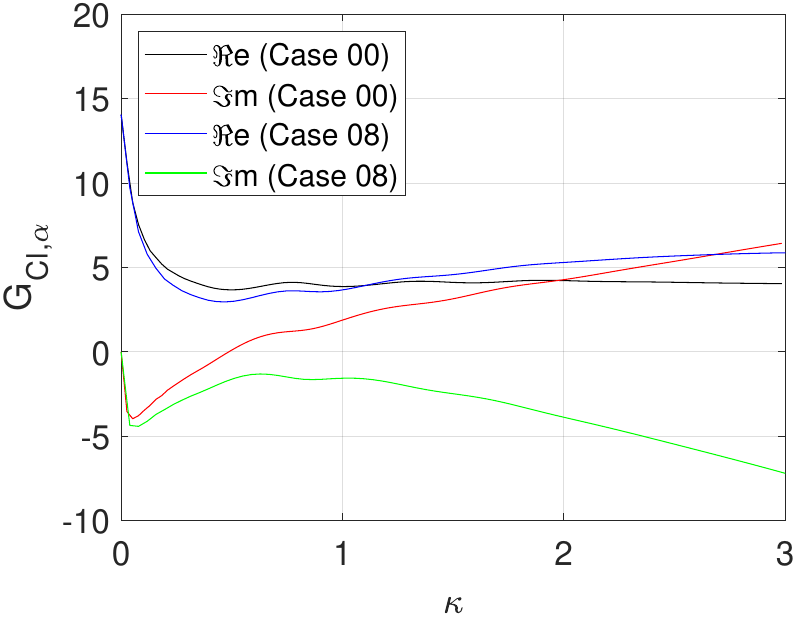}} \\
    	 \subfigure[$G_{C_{m},h}$ results for Case 08 using rectangular window overlaid by Case 00 counterpart.]{ \includegraphics[scale=0.37,trim = 3.5cm 8.5cm 4.5cm 9cm,clip]{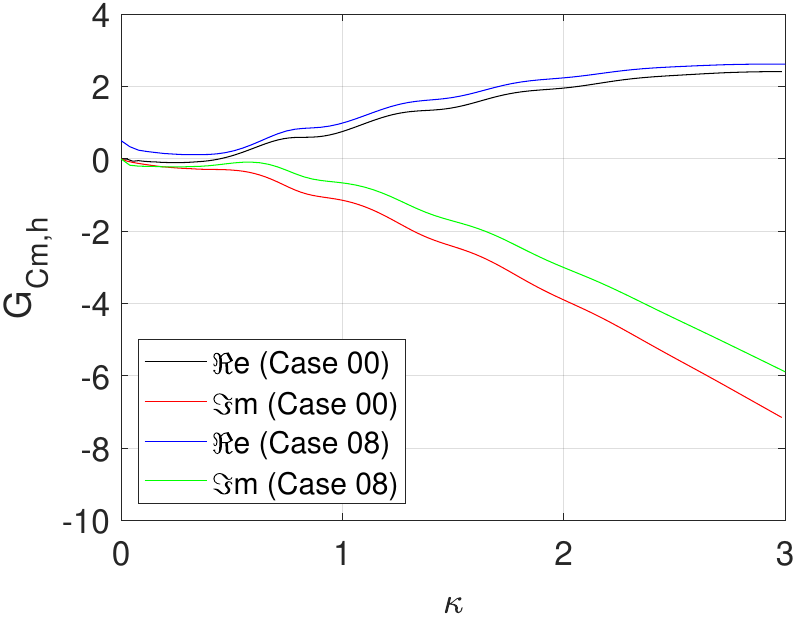}} \quad
    	\subfigure[$G_{C_{m},\alpha}$ results for Case 08 using rectangular window overlaid by Case 00 counterpart.]{ \includegraphics[scale=0.37,trim = 3.5cm 8.5cm 4.5cm 9cm,clip]{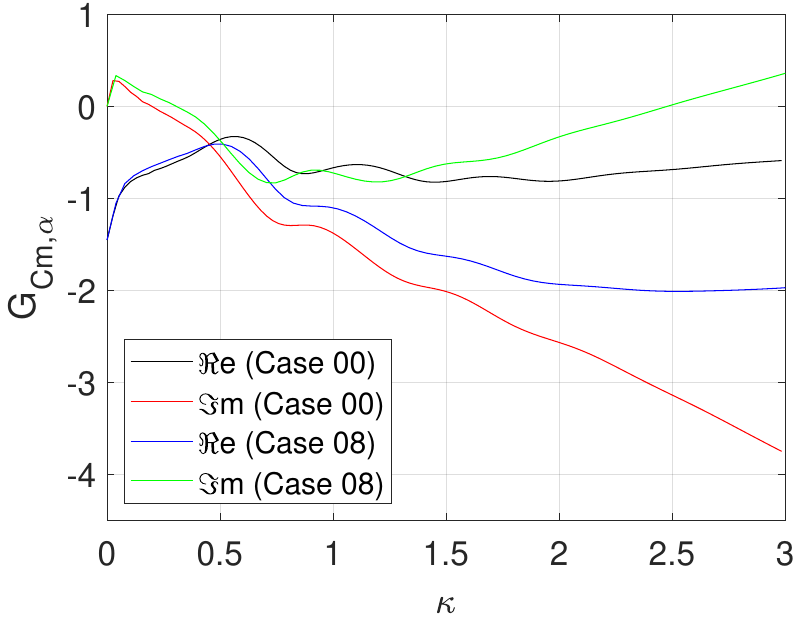}} 
    	 \caption{\label{fig:figure4}Aerodynamic transfer function for Case 08 using rectangular window.}
    	 \vspace{-0.2cm}
    \end{center}
\end{figure}

Transfer functions are obtained for all the 18 test cases but only some of the results are presented in this document, namely the ones that illustrate the potential improvements offered by the signal processing procedure. Initial tests indicate that the prescribed input signals in plunge and pitch modes, and also their derivatives, must be orthogonal to one another in order to apply the simultaneous excitation approach consistently. Such conclusion is based on the assessment of cross-power spectral densities between the input signals. The inputs WF1 and WF3 do not satisfy this linear independence relation either between the input signals or their derivatives. Consequently, all results from these input signals are not capable of reproducing the transfer functions of the benchmarked Case 00 even with the rectangular window. Such behavior can be observed in Figs.\ \ref{fig:figure3} and \ref{fig:figure4} for Cases 02 and 08, for instance.

\begin{figure}[hbt!]
    \begin{center}
    	 \subfigure[$G_{C_{l},h}$ results for Case 11 using rectangular window overlaid by Case 00 counterpart.]{ \includegraphics[scale=0.37,trim = 3.5cm 8.5cm 4.5cm 9cm,clip]{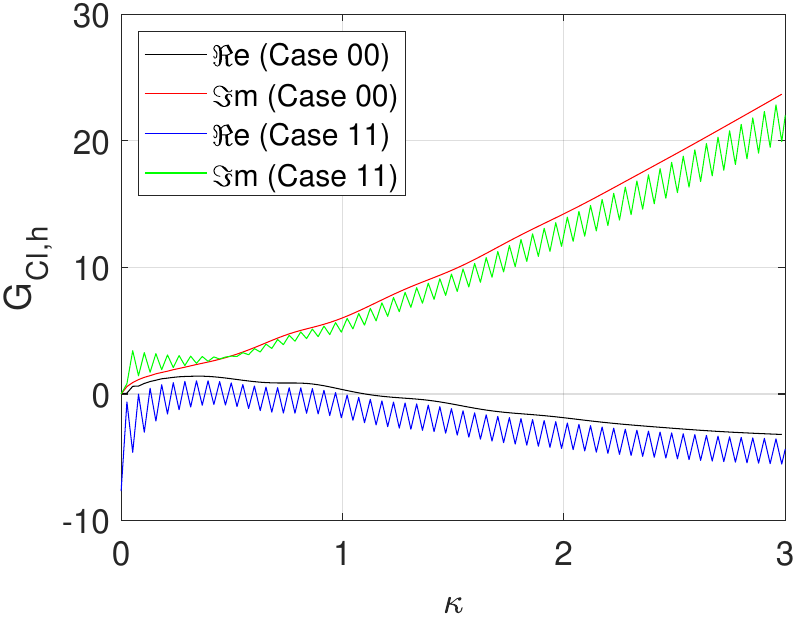}} \quad
    	 \subfigure[$G_{C_{l},\alpha}$ results for Case 11 using rectangular window overlaid by Case 00 counterpart.]{ \includegraphics[scale=0.37,trim = 3.5cm 8.5cm 4.5cm 9cm,clip]{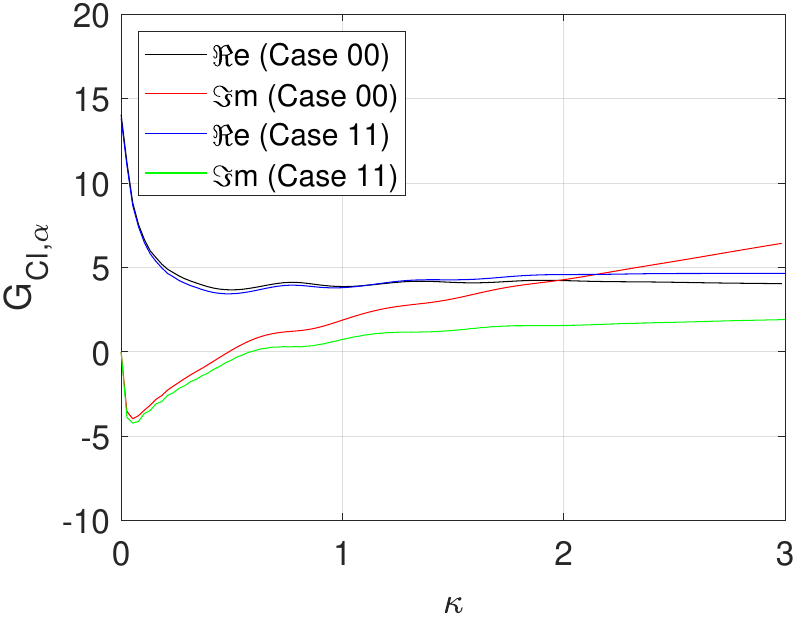}} \\
    	 \subfigure[$G_{C_{m},h}$ results for Case 11 using rectangular window overlaid by Case 00 counterpart.]{ \includegraphics[scale=0.37,trim = 3.5cm 8.5cm 4.5cm 9cm,clip]{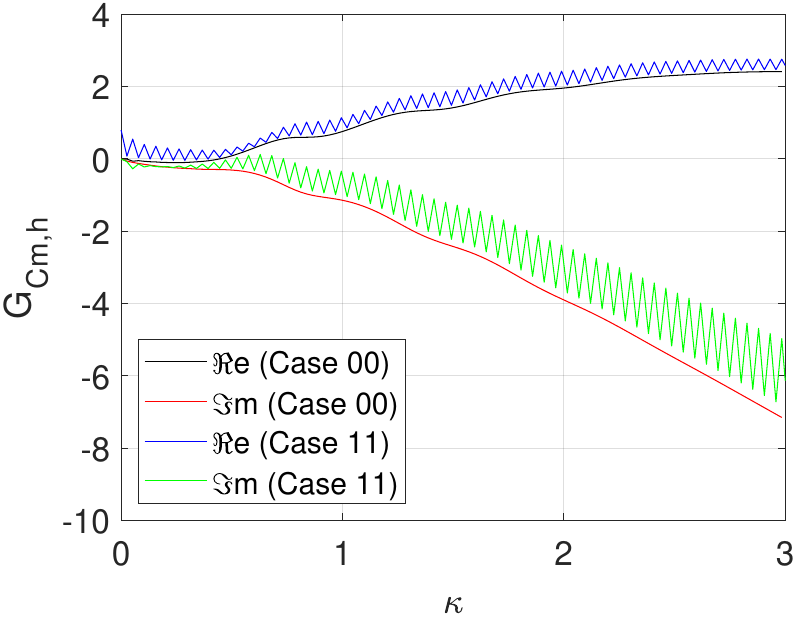}} \quad
    	\subfigure[$G_{C_{m},\alpha}$ results for Case 11 using rectangular window overlaid by Case 00 counterpart.]{ \includegraphics[scale=0.37,trim = 3.5cm 8.5cm 4.5cm 9cm,clip]{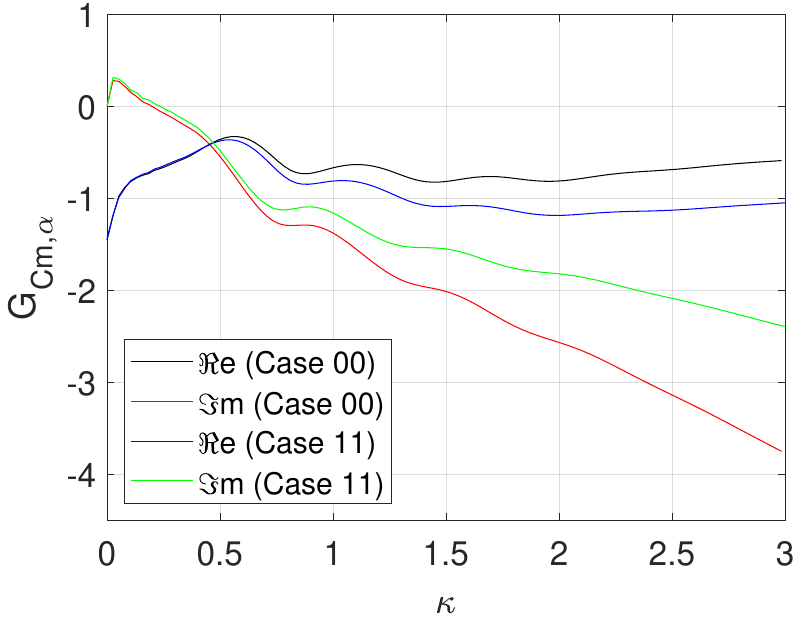}} 
    	 \caption{\label{fig:figure5}Aerodynamic transfer function for Case 11 using rectangular window.}
    	 \vspace{-0.2cm}
    \end{center}
\end{figure}

\begin{figure}[hbt!]
    \begin{center}
    	 \subfigure[$G_{C_{l},h}$ results for Case 11 using Hanning window overlaid by Case 00 counterpart.]{ \includegraphics[scale=0.37,trim = 3.5cm 8.5cm 4.5cm 9cm,clip]{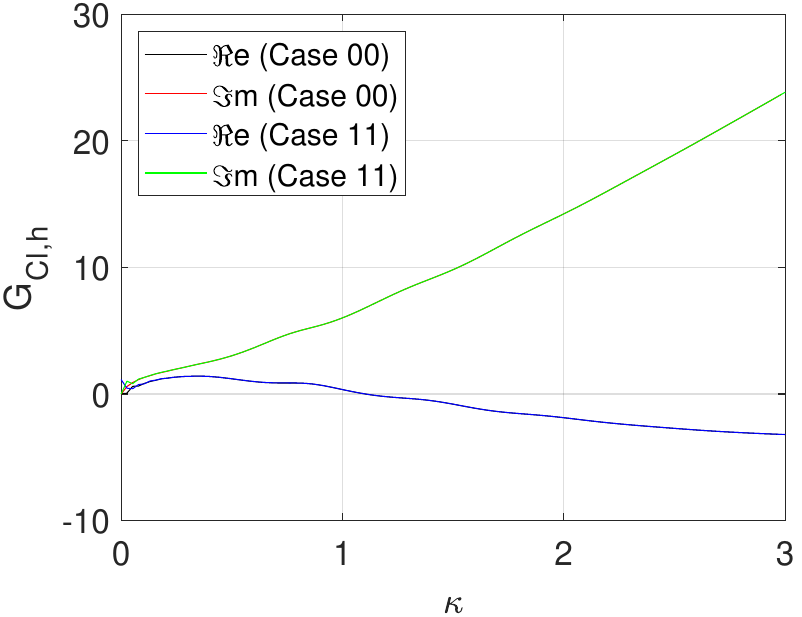}} \quad
    	 \subfigure[$G_{C_{l},\alpha}$ results for Case 11 using Hanning window overlaid by Case 00 counterpart.]{ \includegraphics[scale=0.37,trim = 3.5cm 8.5cm 4.5cm 9cm,clip]{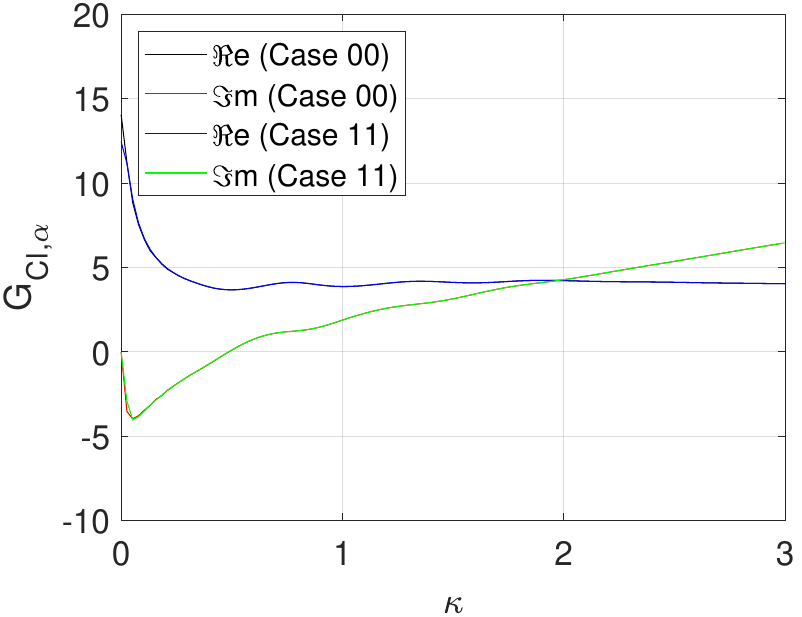}} \\
    	 \subfigure[$G_{C_{m},h}$ results for Case 11 using Hanning window overlaid by Case 00 counterpart.]{ \includegraphics[scale=0.37,trim = 3.5cm 8.5cm 4.5cm 9cm,clip]{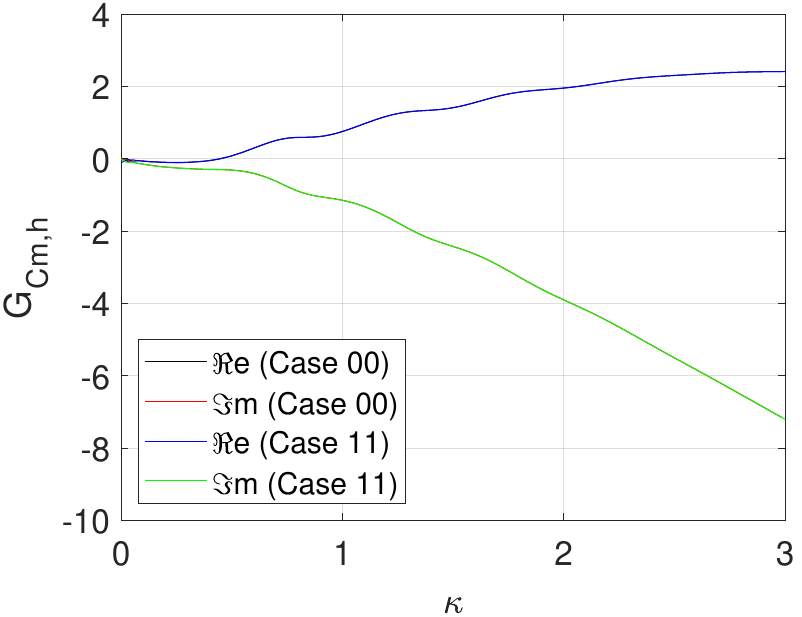}} \quad
    	\subfigure[$G_{C_{m},\alpha}$ results for Case 11 using Hanning window overlaid by Case 00 counterpart.]{ \includegraphics[scale=0.37,trim = 3.5cm 8.5cm 4.5cm 9cm,clip]{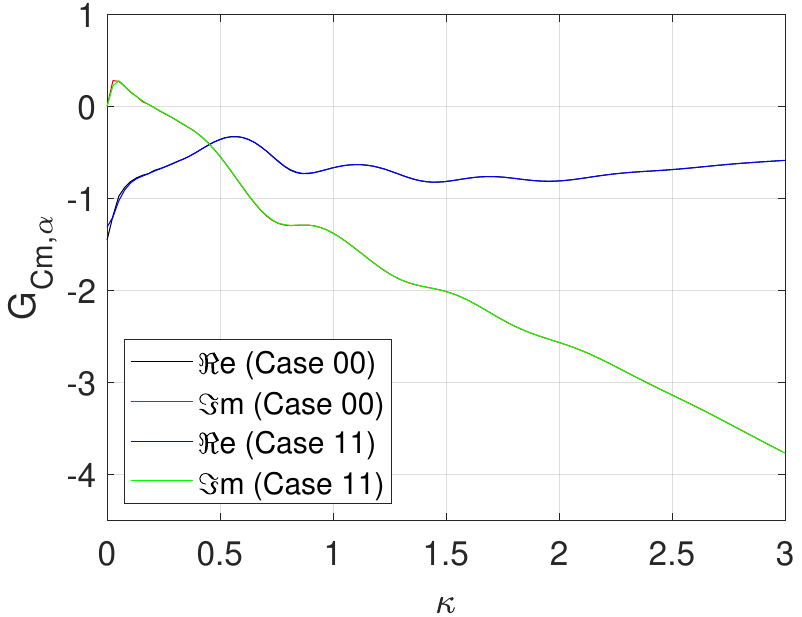}} 
    	 \caption{\label{fig:figure6}Aerodynamic transfer function for Case 11 using Hanning window.}
    	 \vspace{-0.2cm}
    \end{center}
\end{figure}

Figure \ref{fig:figure5} depicts the resulting transfer functions for Case 11 when considering the rectangular window in the signal processing. In agreement with results of Azevedo \emph{et al.} \cite{azevedo2013effects}, one observes that the transfer functions of the plunging mode display a slight spreading but it seems to follow the correct tendency. On the other hand, the transfer functions of the pitching mode are very clear and more precise for lower values of reduced frequencies and tend to diverge as $\kappa$ increases. Conversely, when considering the Hanning window in Case 11, Fig.\ \ref{fig:figure6} shows that the resulting transfer functions for both plunge and pitch modes bear great resemblance to the benchmarked Case 00. Hence, it is clear that the signal processing parameters play a significant role in correctly identifying the aerodynamic transfer functions.

\begin{figure}[hbt!]
    \begin{center}
    	 \subfigure[$G_{C_{l},h}$ results for Case 12 using rectangular window overlaid by Case 00 counterpart.]{ \includegraphics[scale=0.37,trim = 3.5cm 8.5cm 4.5cm 9cm,clip]{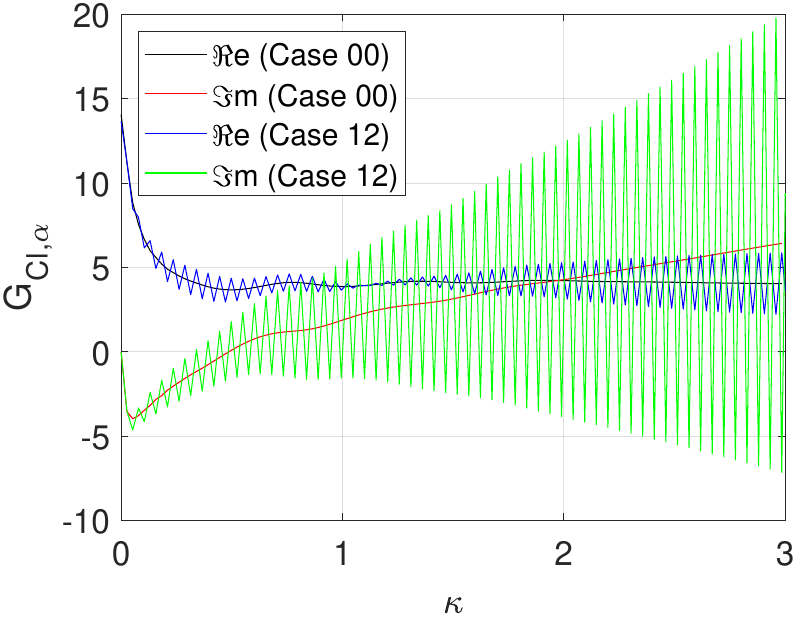}} \quad
    	 \subfigure[$G_{C_{l},\alpha}$ results for Case 12 using rectangular window overlaid by Case 00 counterpart.]{ \includegraphics[scale=0.37,trim = 3.5cm 8.5cm 4.5cm 9cm,clip]{figures/transferFunction/ReIm_G12_caso12.pdf}} \\
    	 \subfigure[$G_{C_{m},h}$ results for Case 12 using rectangular window overlaid by Case 00 counterpart.]{ \includegraphics[scale=0.37,trim = 3.5cm 8.5cm 4.5cm 9cm,clip]{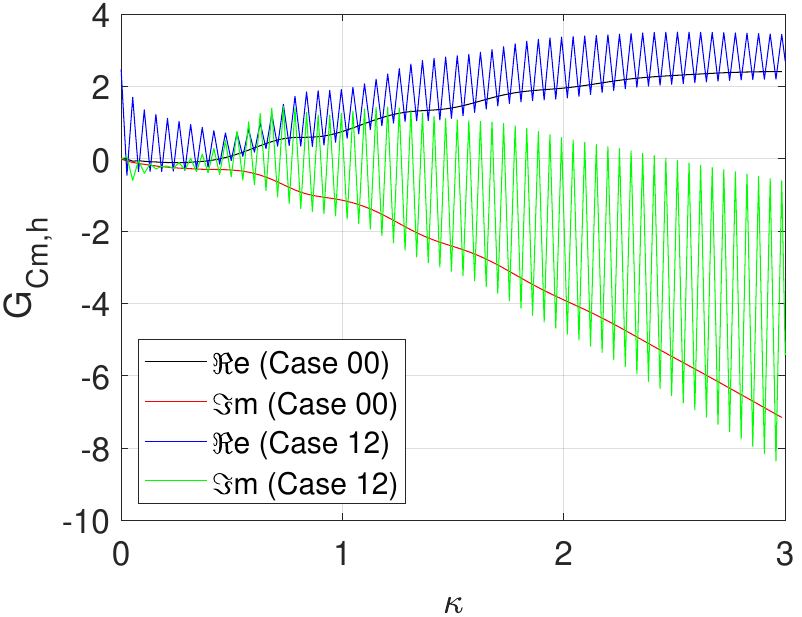}} \quad
    	\subfigure[$G_{C_{m},\alpha}$ results for Case 12 using rectangular window overlaid by Case 00 counterpart.]{ \includegraphics[scale=0.37,trim = 3.5cm 8.5cm 4.5cm 9cm,clip]{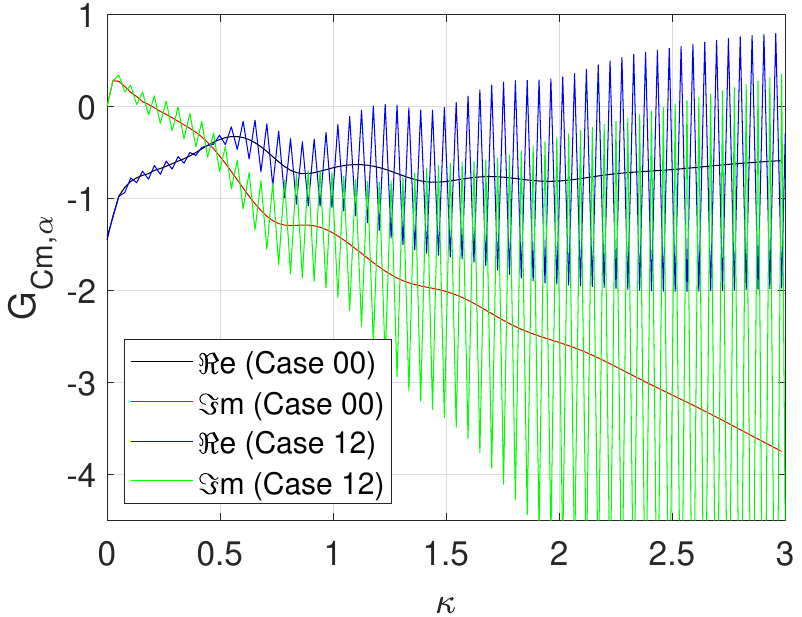}} 
    	 \caption{\label{fig:figure7}Aerodynamic transfer function for Case 12 using rectangular window.}
    	 \vspace{-0.2cm}
    \end{center}
\end{figure}

\begin{figure}[hbt!]
    \begin{center}
    	 \subfigure[$G_{C_{l},h}$ results for Case 12 using Hanning window overlaid by Case 00 counterpart.]{ \includegraphics[scale=0.37,trim = 3.5cm 8.5cm 4.5cm 9cm,clip]{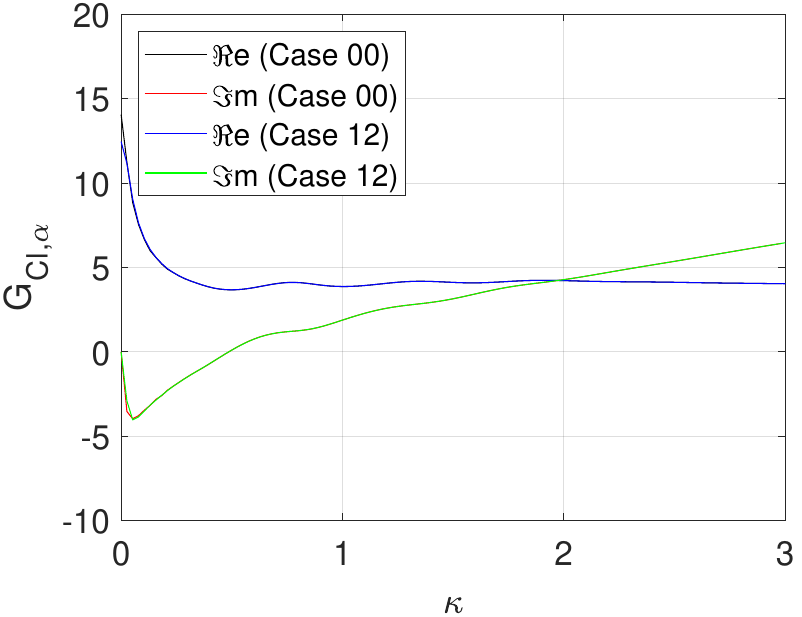}} \quad
    	 \subfigure[$G_{C_{l},\alpha}$ results for Case 12 using Hanning window overlaid by Case 00 counterpart.]{ \includegraphics[scale=0.37,trim = 3.5cm 8.5cm 4.5cm 9cm,clip]{figures/transferFunction/ReIm_G12_caso12_Hanning.pdf}} \\
    	 \subfigure[$G_{C_{m},h}$ results for Case 12 using Hanning window overlaid by Case 00 counterpart.]{ \includegraphics[scale=0.37,trim = 3.5cm 8.5cm 4.5cm 9cm,clip]{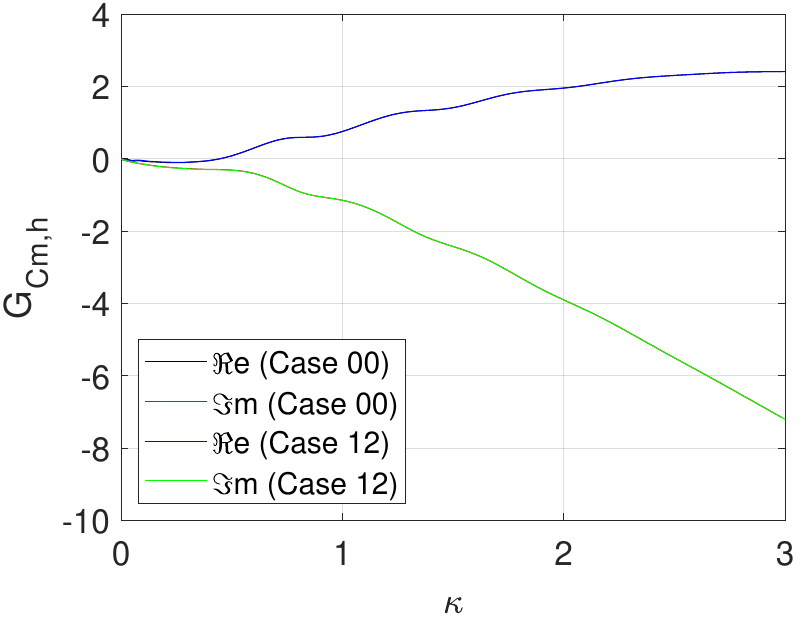}} \quad
    	\subfigure[$G_{C_{m},\alpha}$ results for Case 12 using Hanning window overlaid by Case 00 counterpart.]{ \includegraphics[scale=0.37,trim = 3.5cm 8.5cm 4.5cm 9cm,clip]{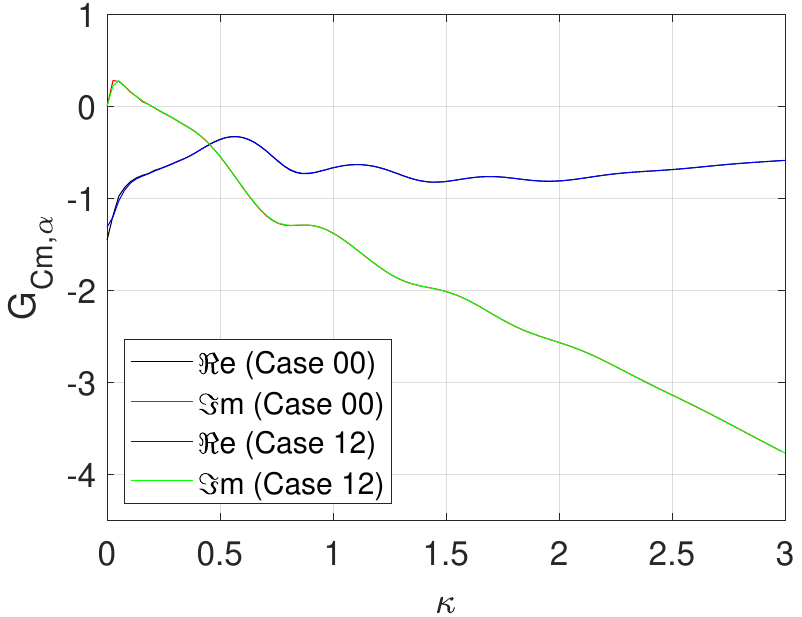}} 
    	 \caption{\label{fig:figure8}Aerodynamic transfer function for Case 12 using Hanning window.}
    	 \vspace{-0.3cm}
    \end{center}
\end{figure}

The transfer functions for Case 12, when using rectangular windows, are very oscillatory for both plunge and pitch modes, as illustrated in Fig.\ \ref{fig:figure7}. In general, this behavior is also present in results from other test sets, which are omitted here for brevity reasons. In view of signal processing techniques, this oscillatory trend is a consequence of the occurrence of the leakage phenomenon. More details about the leakage phenomenon can be found in the work of Oppenheim and Schafer \cite{oppenheim2001discrete}. On the other hand, Fig.\ \ref{fig:figure8} shows that there are no oscillations for both plunge and pitch modes when assessing Case 12 using a Hanning window. The application of the Hanning window in Cases 11 and 12 solves the leakage problem because, by definition \citep{stoica2005spectral}, it cancels the contribution of the first and last point of the original signal. Hence, these results clearly indicate that the usefulness of different test cases in the context of a power spectral density analysis also depends on a suitable treatment of the signal processing aspects of the present approach. Given the fact that both Cases 11 and 12 predict the transfer functions accordingly when applying the Hanning window, these configurations advance through the proposed procedure whereas the cases with rectangular window are not further considered in the present discussion.

\subsection{Polynomial Interpolation}

As mentioned earlier, the final stage in preparing the data for an aeroelastic stability analysis consists in approximating the transfer function values by the selected interpolating polynomial. Here, all calculations considered the first form of the Eversman and Tewari polynomials with the addition of 6 poles. The data fitting is performed for the reduced frequency range of $0 \leq \kappa \leq 3$ because it represents the needed aerodynamic information for the stability analysis in question. Moreover, given the fact that some of the proposed cases failed to produce a coherent set of transfer functions, the authors restricted the cases under evaluation for this stage.

\begin{figure}[hbt!]
    \begin{center}
    	 \subfigure[$G_{C_{l},h}$ results for Case 11 using Hanning window layered over original data set.]{ \includegraphics[scale=0.377,trim = 3.5cm 8.5cm 4.5cm 9cm,clip]{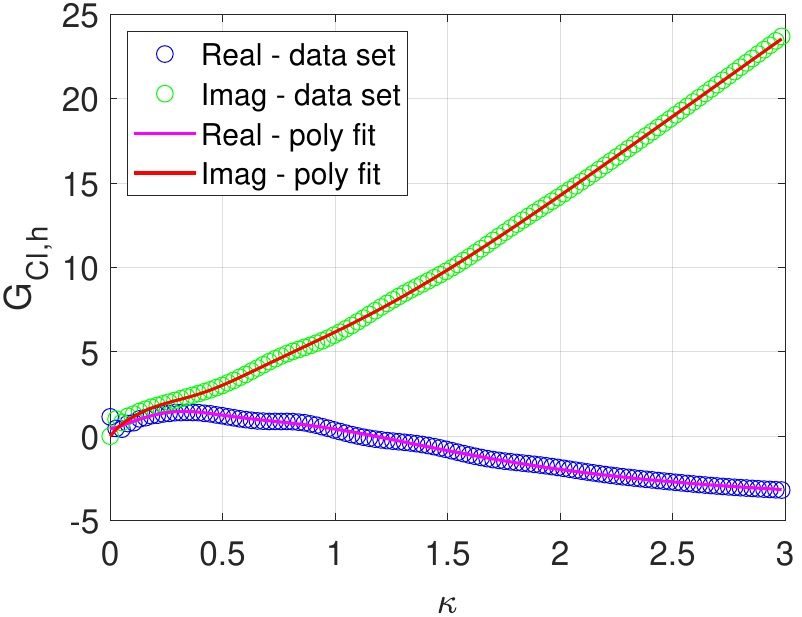}} \quad
    	 \subfigure[$G_{C_{l},\alpha}$ results for Case 11 using Hanning window layered over original data set.]{ \includegraphics[scale=0.377,trim = 3.5cm 8.5cm 4.5cm 9cm,clip]{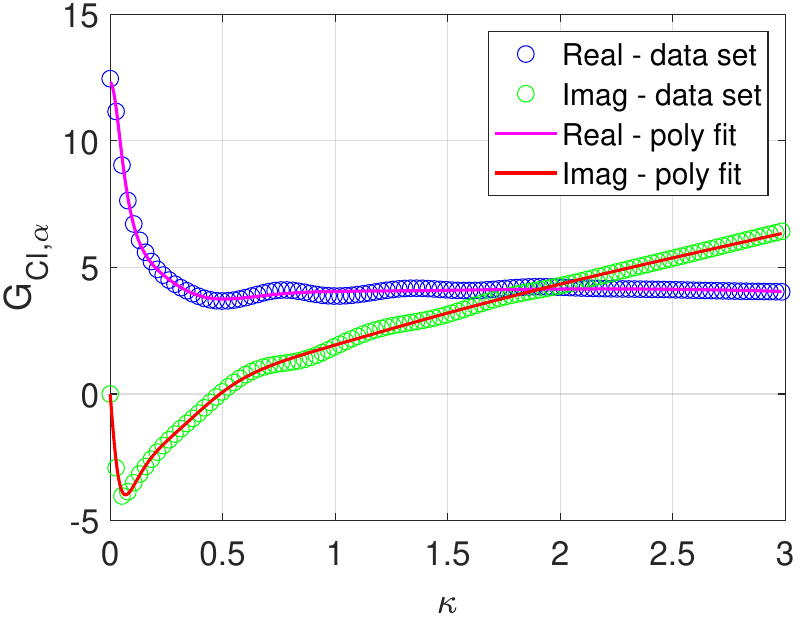}} \\
    	 \subfigure[$G_{C_{m},h}$ results for Case 11 using Hanning window layered over original data set.]{ \includegraphics[scale=0.377,trim = 3.5cm 8.5cm 4.5cm 9cm,clip]{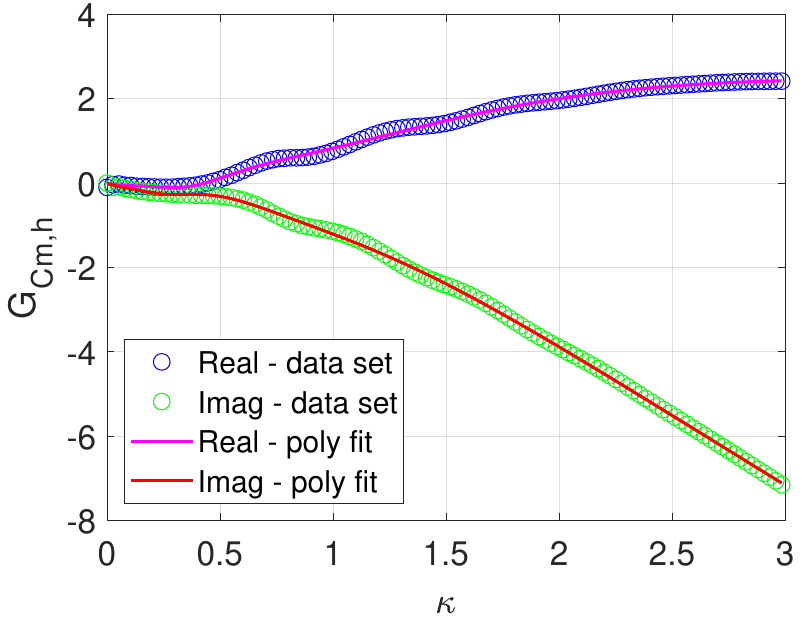}} \quad
    	\subfigure[$G_{C_{m},\alpha}$ results for Case 11 using Hanning window layered over original data set.]{ \includegraphics[scale=0.377,trim = 3.5cm 8.5cm 4.5cm 9cm,clip]{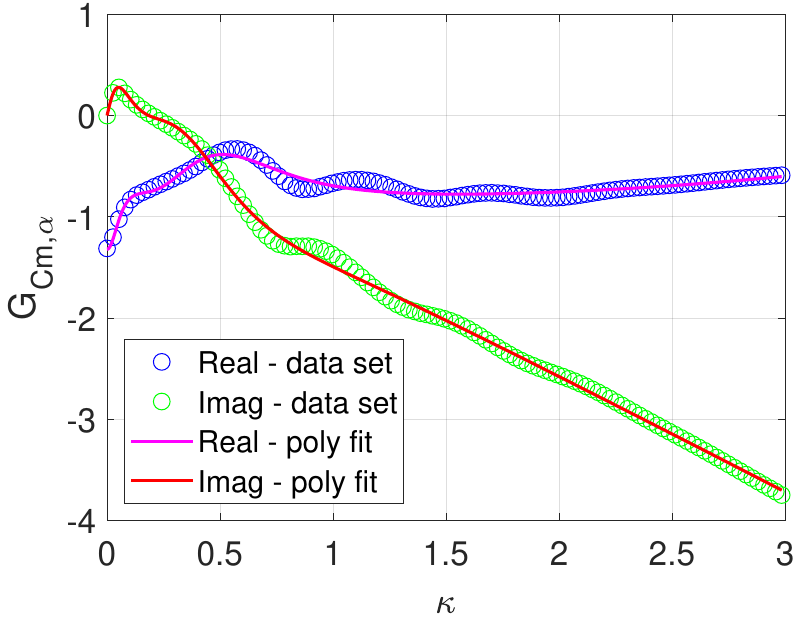}} 
    	 \caption{\label{fig:figure9}Interpolating polynomials for Case 11 using Hanning window.}
    	 \vspace{-0.3cm}
    \end{center}
\end{figure}

\begin{figure}[hbt!]
    \begin{center}
    	 \subfigure[$G_{C_{l},h}$ results for Case 12 using Hanning window layered over original data set.]{ \includegraphics[scale=0.377,trim = 3.5cm 8.5cm 4.5cm 9cm,clip]{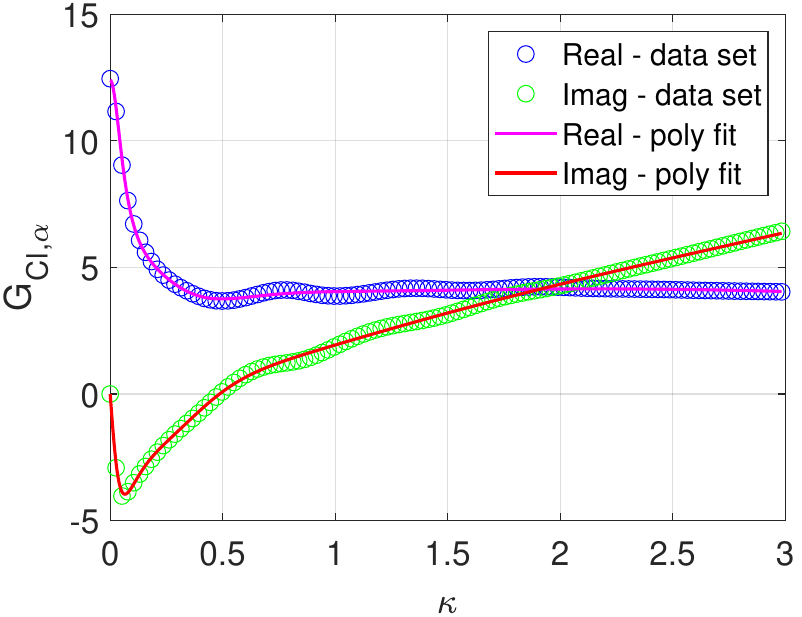}} \quad
    	 \subfigure[$G_{C_{l},\alpha}$ results for Case 12 using Hanning window layered over original data set.]{ \includegraphics[scale=0.377,trim = 3.5cm 8.5cm 4.5cm 9cm,clip]{figures/intpol/IntPolG12_caso12_Hanning.pdf}} \\
    	 \subfigure[$G_{C_{m},h}$ results for Case 12 using Hanning window layered over original data set.]{ \includegraphics[scale=0.377,trim = 3.5cm 8.5cm 4.5cm 9cm,clip]{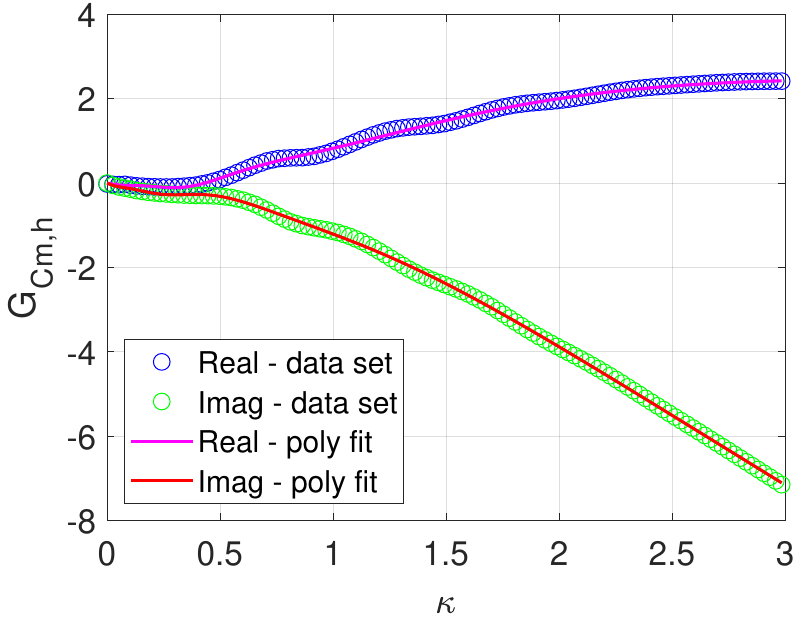}} \quad
    	\subfigure[$G_{C_{m},\alpha}$ results for Case 12 using Hanning window layered over original data set.]{ \includegraphics[scale=0.377,trim = 3.5cm 8.5cm 4.5cm 9cm,clip]{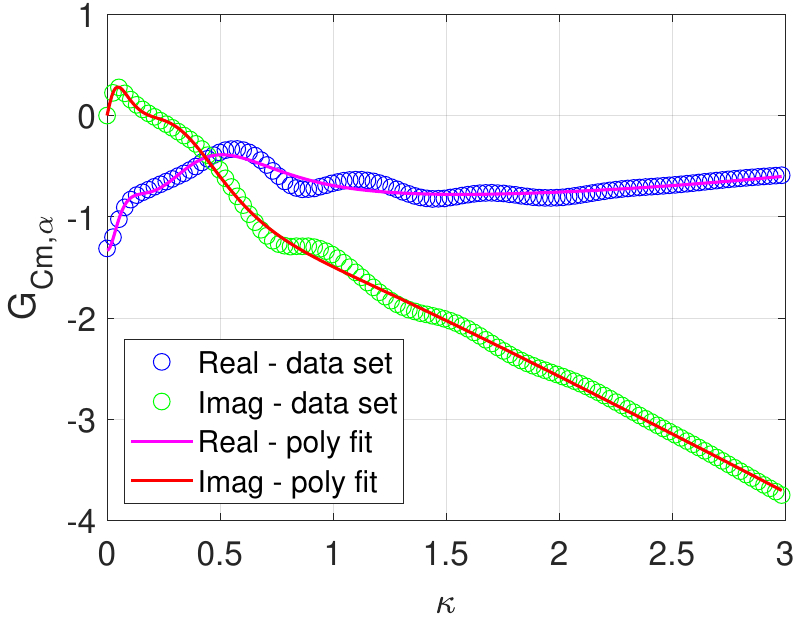}} 
    	 \caption{\label{fig:figure10}Interpolating polynomials for Case 12 using Hanning window.}
    	 \vspace{-0.2cm}
    \end{center}
\end{figure}

Figures \ref{fig:figure9} and \ref{fig:figure10} present the comparison of the interpolated polynomials from the transfer functions of Cases 11 and 12, respectively, over the data from the CFD solution. It should be noticed that the flutter stability analysis is based on data from these approximated transfer functions. As one can see in Figs.\ \ref{fig:figure9} and \ref{fig:figure10}, in general, there is a good agreement between the sets of results from the approximating polynomials and the original CFD data. Discrepancies occur in the extremely low portion of the reduced frequency range analyzed and in regions in which there is a rapid variation in the CFD data. Moreover, the authors have also experienced with the second form of the Eversman and Tewari polynomials. Preliminary results indicate that this second form  indeed produces a better conditioned eigenvalue problem, but it has not improved the correlation with the CFD data. In addition, extrapolating the second form of the Eversman and Tewari polynomials by using a unique pole to describe the approximation is not enough to improve the estimates of the flutter characteristic speed in this application.

\subsection{Aeroelastic Stability Analysis}

\begin{figure}[hbt!]
    \begin{center}
    	 \subfigure[First structural mode for Case 11 using Hanning window.]{ \includegraphics[scale=0.417,trim = 3.5cm 8cm 4.5cm 9cm,clip]{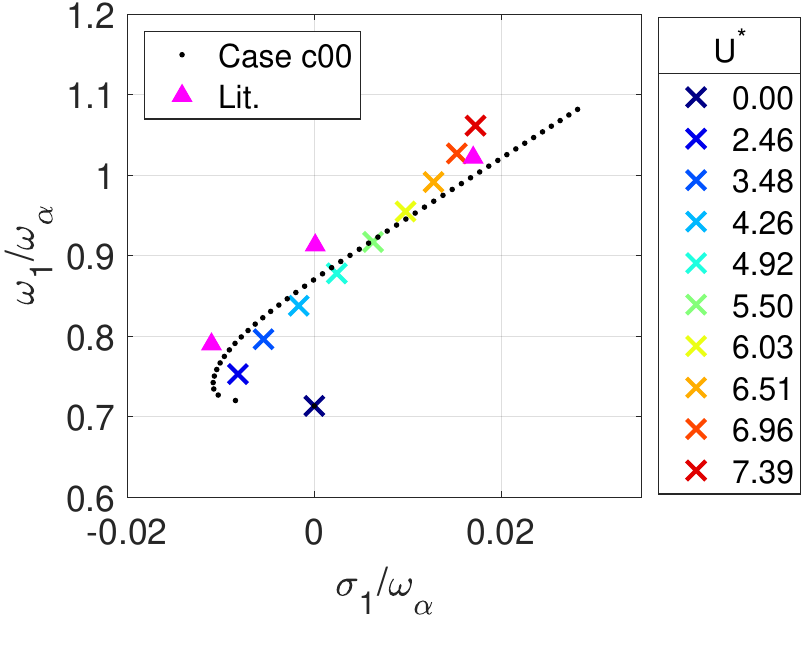}} \quad
    	 \subfigure[Second structural mode for Case 11 using Hanning window.]{ \includegraphics[scale=0.417,trim = 3.5cm 8cm 4.5cm 9cm,clip]{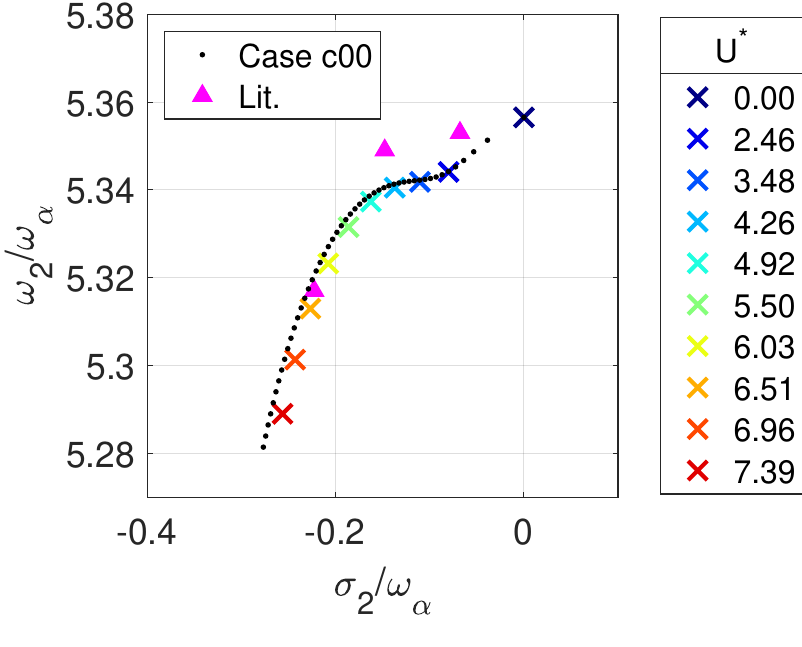}} 
    	 \caption{\label{fig:figure11}Root locus of the first and second structural modes for Case 11 using Hanning window.}
    	 \vspace{-0.2cm}
    \end{center}
\end{figure}

\begin{figure}[hbt!]
    \begin{center}
    	 \subfigure[First structural mode for Case 12 using Hanning window.]{ \includegraphics[scale=0.417,trim = 3.5cm 8cm 4.5cm 9cm,clip]{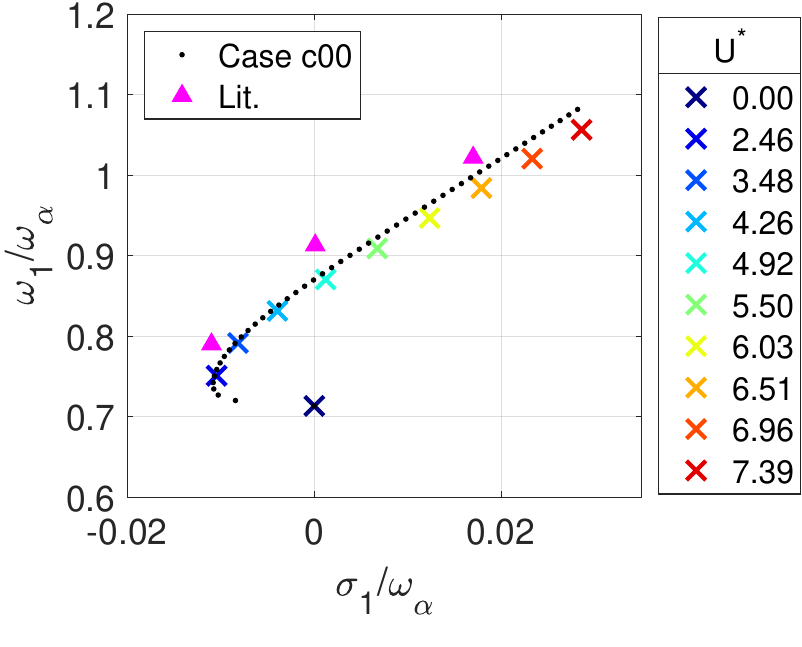}} \quad
    	 \subfigure[Second structural mode for Case 12 using Hanning window.]{ \includegraphics[scale=0.417,trim = 3.5cm 8cm 4.5cm 9cm,clip]{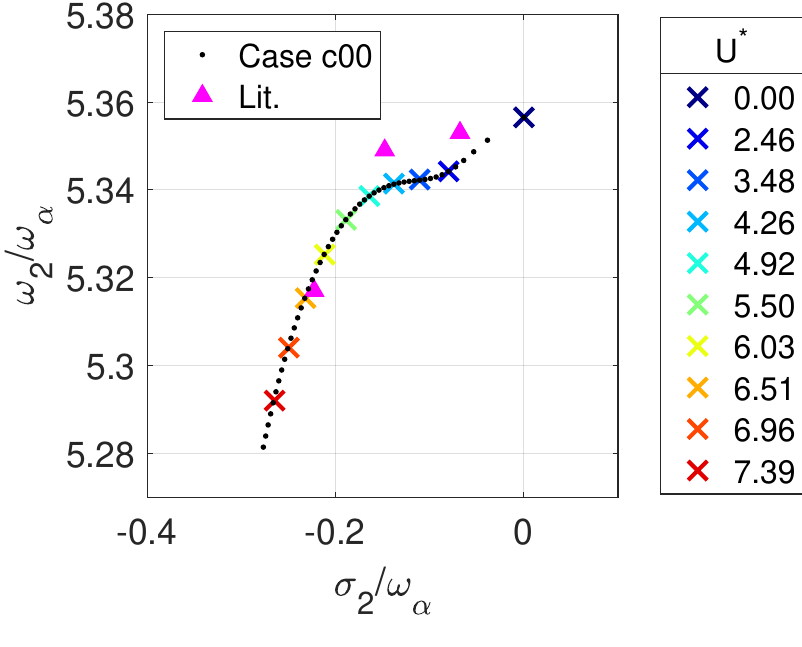}} 
    	 \caption{\label{fig:figure12}Root locus of the first and second structural modes for Case 12 using Hanning window.}
    	 \vspace{-0.2cm}
    \end{center}
\end{figure}

Finally, the obtained polynomials are employed in the solution of the flutter stability eigenvalue problem. Some stability root loci of the first and second structural modes can be seen in Figs.\ \ref{fig:figure11} and \ref{fig:figure12} for Cases 11 and 12, respectively. Both figures present comparisons with numerical results from Case 00 and from Rausch \emph{et al.} \cite{rausch1990euler}, the latter referred to as ``Lit.''. The characteristic dynamic pressure parameter varies from $Q^{*}=0.0$ up to 1.0 in $\Delta Q^{*}=0.01$ intervals. Each plotted point of characteristic speed corresponds to one of these values. The literature data, however, are only available for $Q^{*}=$ 0.2, 0.5, and 0.8\@. Figures \ref{fig:figure11} and \ref{fig:figure12} show that the modal damping and frequency behavior of aeroelastic modes predicted by the proposed procedure using discrete step and Walsh functions are very close. Results from Case 12 clearly show even more similarities in comparison with the reference data than the Case 11 counterpart. An important aspect to be considered in the root locus analysis is that one is concerned with the flutter onset point, \textit{i.e.}, the flutter characteristic speed. It means that the characteristic speed associated with flutter phenomenon is as relevant as the accuracy of all data in the root locus plot. Rausch \emph{et al.} \cite{rausch1990euler} state that flutter occurs close to $Q^{*}=0.5$, which is a near neutrally stable condition for the same aeroelastic configuration considered herein. It corresponds to a reference characteristic speed of $U^{*}=5.4772$. 

Considering these literature data as benchmark, Table \ref{tab:table2} presents the modal damping and frequency values at the flutter point for the test cases capable of reproducing the transfer functions with either rectangular or Hanning window. By definition, the flutter onset point can be identified as the point at which the damping of one of the aeroelastic modes becomes identically zero. Results displayed in Table \ref{tab:table2}, however, present the first point in which the associated eigenvalue has a positive imaginary part. Therefore, it does not necessarily happen when the damping is identically zero. Given the fact that $\Delta Q^{*}=0.01$, the identified results for the flutter onset point are considered to be sufficiently accurate. The percentage error indicated in Table \ref{tab:table2} compares the flutter characteristic speed, $U^{*}_{f}$, with the aforementioned reference data. One can observe that Cases 05, 13, and 18 predict the flutter onset point with percentage errors lower than 5\%, whereas the remaining cases present a percentage error within the range 8.9\% to 15.9\%. In general, all results are conservative since they are below the reference value.

\begin{table}[hbt!]
\caption{\label{tab:table2} Flutter points for NACA 0012 airfoil at $M_{\infty}$=0.80 for different test cases.} 
    \begin{center} {
		\begin{tabular}{ccccccccc}
		\hline
        Input & Case & Window & $\sigma_1/\omega_\alpha$ & $\omega_1/\omega_\alpha$ & $\sigma_2/\omega_\alpha$ & $\omega_2/\omega_\alpha$ & $U^{*}_{f}$ & Error \%  \\ \hline
        Step  & 00 & Rectangular & 3.5660$\times$10$^{-6}$ & 0.8735  & -0.1669 & 5.3383  & 4.9848 & 9.0 \\
        WF2   & 04 & Rectangular & 4.9512$\times$10$^{-4}$ & 0.8713  & -0.1677 & 5.3386  & 4.9848 & 8.9 \\
        WF2   & 05 & Rectangular & 2.5028$\times$10$^{-4}$ & 0.8761  & -0.1821 & 5.3397  & 5.2223 & 4.6 \\
        WF4   & 11 & Hanning & 3.7337$\times$10$^{-4}$ & 0.8581  & -0.1502 & 5.3392 & 4.6057 & 15.9 \\
        WF4   & 12 & Hanning & 1.3968$\times$10$^{-4}$ & 0.8627  & -0.1592 & 5.3394 & 4.7990 & 12.4  \\
        WF4   & 13 & Rectangular & 2.8385$\times$10$^{-4}$ & 0.8958  & -0.1840 & 5.3338  & 5.3936 & 1.5 \\
        WF5   & 16 & Hanning & 1.3133$\times$10$^{-5}$   & 0.8667 & -0.1647 & 5.3385 & 4.9237 & 10.1 \\
        WF5   & 17 & Hanning & 2.4130$\times$10$^{-5}$   & 0.8654 & -0.1638 & 5.3379 & 4.9237 & 10.1 \\
        WF5   & 18 & Rectangular & 3.1175$\times$10$^{-5}$ & 0.8887  & -0.1764 & 5.3358  & 5.2223 & 4.7 \\
        \hline
	    \end{tabular}}
	    \vspace{-0.5cm}
    \end{center}
\end{table}

\section{Concluding Remarks}

The authors present a reduced-order model capable of modeling unsteady transonic aerodynamic loads using system identification techniques. Results are presented for a NACA 0012 airfoil-based typical section, for freestream conditions $M_{\infty}=0.8$ and $\alpha_{0}=0$. From the steady-solution, the aeroelastic system is perturbed in a prescribed pattern in order to obtain unsteady CFD-based aerodynamic results. The prescribed airfoil movements are defined as step inputs and Walsh functions respectively representing the mode-by-mode and simultaneous excitation approaches. Initially, it is shown that the applied non-parametric system identification technique is capable of splitting the output into the contribution of each mode to the corresponding aerodynamic transfer functions. Results show that an accurate identification of the aerodynamic transfer functions depends on two aspects, namely the orthogonality of the input signals and their derivatives to one another, and also a proper signal processing. Afterwards, it is shown that the discrete strings of transfer functions can be conveniently approximated using interpolating polynomials. Such polynomials contain, therefore, the aerodynamic states, which can be used for aeroelastic stability analyses in the frequency domain. Finally, these rational-function approximations are used to generate root locus plots for the aeroelastic system that, in general, ultimately results in percentage errors lower than 16\% between the computed and the reference characteristic speed values.  

\section{Acknowledgements}
The authors gratefully acknowledge the support for this research provided by Fundação de Amparo à Pesquisa do Estado de São Paulo, FAPESP, through a Master of Science Scholarship for the first author according to FAPESP Process No.\ 2022/01397-0\@. This work was also supported by Fundação Coordenação de Aperfeiçoamento de Pessoal de Nível Superior, CAPES, under the Research Grants No.\ 88887.634461/2021-00 and 88887.609895/2021-00\@. Additional support provided by FAPESP under the Research Grant No.\ 2013/07375-0 is also gratefully acknowledged. The present work has also received support from Conselho Nacional de Desenvolvimento Cient\'{\i}fico e Tecnol\'{o}gico, CNPq, under the Research Grant No.\ 309985/2013-7\@.

\bibliographystyle{plain}
\bibliography{biblio}

\begin{thebibliography}{10}

\bibitem{azevedo2012efficient}
João Henrique~F. Azevedo, João Luiz~F. Azevedo, and Roberto G.~A. Silva.
\newblock {Efficient Calculation of Aerodynamic States for Aeroelastic Analyses
  in the Frequency Domain}.
\newblock AIAA Paper No.\ 2012-1716, \emph{53rd AIAA/ASME/ASCE/AHS/ASC
  Structures, Structural Dynamics and Materials Conference}, Honolulu, HI, Apr.
  2012.

\bibitem{azevedo2013effects}
João Henrique~F. Azevedo, João Luiz~F. Azevedo, and Roberto G.~A. Silva.
\newblock {Effects of the Aerodynamic Data in a {MIMO} System Identification
  Framework for Aeroelastic Analyses}.
\newblock AIAA Paper No.\ 2013-1486, \emph{54th AIAA/ASME/ASCE/AHS/ASC
  Structures, Structural Dynamics, and Materials Conference}, Boston, MA, Apr.
  2013.

\bibitem{bisplinghoff1955aeroelasticity}
R.~L. Bisplinghoff, H.~Ashley, and R.~L. Halfman.
\newblock {\em {Aeroelasticity}}.
\newblock Addison-Wesley Publishing Company, Inc., 1955.

\bibitem{camilo2013hopf}
Elizangela Camilo, Flávio~D. Marques, and João Luiz~F. Azevedo.
\newblock {Hopf Bifurcation Analysis of Typical Sections with Structural
  Nonlinearities in Transonic Flow}.
\newblock {\em Aerospace Science and Technology}, 30(1):163--174, Oct. 2013.

\bibitem{eversman1991consistent}
Walter Eversman and Ashish Tewari.
\newblock {Consistent Rational-Function Approximation for Unsteady
  Aerodynamics}.
\newblock {\em Journal of Aircraft}, 28(9):545--552, 1991.

\bibitem{hadamard93}
Jacques Hadamard.
\newblock {R\`{e}solution d'une Question Relative aux D\`{e}terminants}.
\newblock {\em Bulletin des Sciences Math\'{e}matiques}, 2(17):240--246, 1893.

\bibitem{marques2008z}
Alexandre~N. Marques and João Luiz~F. Azevedo.
\newblock {A Z-Transform Discrete-Time State-Space Formulation for Aeroelastic
  Stability Analysis}.
\newblock {\em Journal of Aircraft}, 45(5):1564--1578, Sept.--Oct. 2008.

\bibitem{marques2008numerical}
Alexandre~N. Marques and João Luiz~F. Azevedo.
\newblock {Numerical Calculation of Impulsive and Indicial Aerodynamic
  Responses Using Computational Aerodynamics Techniques}.
\newblock {\em Journal of Aircraft}, 45(4):1112--1135, July--Aug. 2008.

\bibitem{oliveira1993metodologia}
Luiz~Cláudio Oliveira.
\newblock {Uma Metodologia de Análise Aeroelástica com Variáveis de Estado
  Utilizando Técnicas de Ae\-ro\-di\-n\^{a}\-mi\-ca Computacional}.
\newblock Master's thesis, Instituto Tecnológico de Aeronáutica, São José
  dos Campos, Brazil, 1993.

\bibitem{oppenheim2001discrete}
Alan~V. Oppenheim and Ronald~W. Schafer.
\newblock {\em {Discrete-Time Signal Processing}}.
\newblock Pearson Prentice Hall, Upper Saddle River, NJ, 2010.

\bibitem{rausch1990euler}
Russ~D. Rausch, John~T. Batina, and Henry T.~Y. Yang.
\newblock {Euler Flutter Analysis of Airfoils Using Unstructured Dynamic
  Meshes}.
\newblock {\em Journal of Aircraft}, 27(5):436--443, 1990.

\bibitem{silva2004development}
Walter Silva and Robert~E. Bartels.
\newblock {Development of Reduced-Order Models for Aeroelastic Analysis and
  Flutter Prediction Using the {CFL}3{D}v6.0 Code}.
\newblock {\em Journal of Fluids and Structures}, 19(6):729--745, July 2004.

\bibitem{silva2008simultaneous}
Walter~A. Silva.
\newblock {Simultaneous Excitation of Multiple-Input/Multiple-Output
  {CFD}-Based Unsteady Aerodynamic Systems}.
\newblock {\em Journal of Aircraft}, 45(4):1267--1274, July--Aug. 2008.

\bibitem{silva2014evaluation}
Walter~A. Silva, Pawel Chwalowski, and Boyd Perry.
\newblock {Evaluation of Linear, Inviscid, Viscous, and Reduced-Order Modelling
  Aeroelastic Solutions of the {AGARD} 445.6 Wing Using Root Locus Analysis}.
\newblock {\em International Journal of Computational Fluid Dynamics},
  28(3-4):122--139, Aug. 2014.

\bibitem{silva2009development}
Walter~A. Silva, Veer~N. Vatsa, and Robert~T. Biedron.
\newblock {Development of Unsteady Aerodynamic and Aeroelastic Reduced-Order
  Models Using the {FUN}3{D} Code}.
\newblock {\em IFASD Paper 2009--030}, June 2009.

\bibitem{skujins2014reduced}
Torstens Skujins and Carlos E.~S. Cesnik.
\newblock {Reduced-Order Modeling of Unsteady Aerodynamics Across Multiple
  {M}ach Regimes}.
\newblock {\em Journal of Aircraft}, 51(6):1681--1704, Nov.--Dec. 2014.

\bibitem{stoica2005spectral}
Petre Stoica, Randolph~L Moses, et~al.
\newblock {\em {Spectral Analysis of Signals}}, volume 452.
\newblock Pearson Prentice Hall, Upper Saddle River, NJ, 2005.

\bibitem{tiffany1987nonlinear}
Sherwood~H. Tiffany and William~M. Adams.
\newblock {Nonlinear Programming Extensions to Rational Function Approximations
  of Unsteady Aerodynamics}.
\newblock Technical Report NASA--TP--2776, NASA, July 1988.

\bibitem{waite2019reduced}
Josiah Waite, Bret Stanford, Robert~E. Bartels, and Walter~A. Silva.
\newblock {Reduced Order Modeling for Transonic Aeroservoelastic Control Law
  Development}.
\newblock pages 1--16, AIAA Paper No.\ 2019-1022, \emph{AIAA Scitech Forum
  2019}, San Diego, CA, Jan. 2019.

\bibitem{wright2008introduction}
J.~R. Wright and J.~E. Cooper.
\newblock {\em {Introduction to Aircraft Aeroelasticity and Loads}}, volume~20.
\newblock John Wiley \& Sons, 2008.

\end{thebibliography}

\end{document}